\documentclass[12pt,manuscript]{aastex}

\newcommand{\kms}{\ensuremath{\rm km\,s^{-1}}}
\newcommand{\ms}{\ensuremath{\rm m\,s^{-1}}}
\newcommand{\gcmc}{\ensuremath{\rm g\,cm^{-3}}}

\newcommand{\teff}{\ensuremath{T_{\rm eff}}}
\newcommand{\logg}{\ensuremath{\log{g}}}
\newcommand{\vsini}{\ensuremath{v \sin{i}}}
\newcommand{\feh}{[Fe/H]}

\newcommand{\rsun}{\ensuremath{R_\sun}}
\newcommand{\msun}{\ensuremath{M_\sun}}
\newcommand{\lsun}{\ensuremath{L_\sun}}

\newcommand{\rstar}{\ensuremath{R_\star}}
\newcommand{\mstar}{\ensuremath{M_\star}}
\newcommand{\loggstar}{\ensuremath{\logg_\star}}
\newcommand{\lstar}{\ensuremath{L_\star}}

\newcommand{\rhostar}{\ensuremath{\rho_\star}}

\newcommand{\rpl}{\ensuremath{R_{\rm P}}}
\newcommand{\mpl}{\ensuremath{M_{\rm P}}}

\newcommand{\rhopl}{\ensuremath{\rho_{\rm P}}}
\newcommand{\loggpl}{\ensuremath{\logg_{\rm P}}}
\newcommand{\teq}{\ensuremath{T_{\rm eq}}}

\newcommand{\rjup}{\ensuremath{R_{\rm J}}}
\newcommand{\mjup}{\ensuremath{M_{\rm J}}}

\newcommand{\rearth}{\ensuremath{R_{\earth}}}
\newcommand{\mearth}{\ensuremath{M_{\earth}}}

\def\note #1]{{\bf #1]}}

\newcommand{\koi}{KOI-72~}
\newcommand{\koib}{KOI-72.01}
\newcommand{\koic}{KOI-72.02}
\newcommand{\starname}{Kepler-10}
\newcommand{\planetb}{Kepler-10b}

\newcommand{\kicid}{KIC~11904151}
\newcommand{\tmid}{2MASS 19024305+5014286}

\newcommand{\kicra}{\ensuremath{19^{\mathrm{h}}02^{\mathrm{m}}43^{\mathrm{s}}.05}}
\newcommand{\kicdec}{\ensuremath{+50^{\circ}14'28''.68}}
\newcommand{\kepmag}{10.96}
\newcommand{\kicteff}{5491}
\newcommand{\kiclogg}{4.47}
\newcommand{\kicradius}{0.983}

\newcommand{\teffReconOrig}{\ensuremath{5750 \pm 125}}     
\newcommand{\loggReconOrig}{\ensuremath{4.5 \pm 0.25}}   
\newcommand{\vsiniReconOrig}{\ensuremath{0.0^{+2}_{-0}}}
\newcommand{\teffRecon}{\ensuremath{5680 \pm 91}}
\newcommand{\loggRecon}{\ensuremath{4.33 \pm 0.16}}
\newcommand{\vsiniRecon}{\ensuremath{1.5 \pm 0.5}}
\newcommand{\fehRecon}{\ensuremath{-0.09 \pm 0.04}}

\newcommand{\teffSMEOrig}{\ensuremath{5705 \pm 150}}    
\newcommand{\fehSMEOrig}{\ensuremath{-0.15 \pm 0.03}}   
\newcommand{\loggSMEOrig}{\ensuremath{4.54 \pm 0.10}}   
\newcommand{\vsiniSMEOrig}{\ensuremath{0.5 \pm 0.5}}    

\newcommand{\teffSME}{\ensuremath{5627 \pm 44}}     

\newcommand{\fehSME}{\ensuremath{-0.15 \pm 0.04}}   
\newcommand{\loggSME}{\ensuremath{4.35 \pm 0.06}}   
\newcommand{\vsiniSME}{\ensuremath{0.5 \pm 0.5}}     

\newcommand{\mstarKASC}{\ensuremath{0.895 \pm 0.060}}
\newcommand{\rstarKASC}{\ensuremath{1.056 \pm 0.021}}
\newcommand{\rhostarKASC}{\ensuremath{1.068 \pm 0.008}}
\newcommand{\loggKASC}{\ensuremath{4.341 \pm 0.012}}
\newcommand{\lumKASC}{\ensuremath{1.004 \pm 0.059}}
\newcommand{\absMag}{\ensuremath{4.746 \pm 0.063}}
\newcommand{\ageKASC}{\ensuremath{11.9 \pm 4.5}}
\newcommand{\distance}{\ensuremath{173 \pm 27}}

\newcommand{\periodb}{\ensuremath{0.837495^{+0.000004}_{-0.000005}}}
\newcommand{\epochb}{\ensuremath{2454964.57375^{+0.00060}_{-0.00082}}}
\newcommand{\depthb}{\ensuremath{152 \pm 4}}
\newcommand{\durationb}{\ensuremath{1.811 \pm 0.024}}
\newcommand{\phaseAmpb}{\ensuremath{7.6 \pm 2.0}}
\newcommand{\occultationb}{\ensuremath{5.8 \pm 2.5}}
\newcommand{\scaledSemiMajb}{\ensuremath{3.436^{+0.070}_{-0.092}}}
\newcommand{\scaledPlanetRadiusb}{\ensuremath{0.01232^{+0.00013}_{-0.00016}}}
\newcommand{\impactb}{\ensuremath{0.339^{+0.073}_{-0.079}}}
\newcommand{\inclinationb}{\ensuremath{84.4\fdg0^{+1.1}_{-1.6}}}

\newcommand{\periodc}{\ensuremath{45.29485^{+0.00065}_{-0.00076}}}
\newcommand{\epochc}{\ensuremath{2454971.6761^{+0.0020}_{-0.0023}}}
\newcommand{\depthc}{\ensuremath{376 \pm 9}}
\newcommand{\durationc}{\ensuremath{6.86 \pm 0.07}}

\newcommand{\scaledSemiMajc}{\ensuremath{49.1^{+1.2}_{-1.3}}}
\newcommand{\scaledPlanetRadiusc}{\ensuremath{0.019378^{+0.00020}_{-0.00024}}}
\newcommand{\impactc}{\ensuremath{0.299^{+0.089}_{-0.073}}}
\newcommand{\inclinationc}{\ensuremath{89.7\fdg0^{+0.09}_{-0.12}}}

\newcommand{\semiAmpb}{\ensuremath{3.3^{+0.8}_{-1.0}}}
\newcommand{\eccb}{\ensuremath{0}}
\newcommand{\gammaVelb}{\ensuremath{-98.93\pm0.02}}

\newcommand{\mplanetb}{\ensuremath{4.56^{+1.17}_{-1.29}}}
\newcommand{\rplanetb}{\ensuremath{1.416^{+0.033}_{-0.036}}}
\newcommand{\rhoplanetb}{\ensuremath{8.8^{+2.1}_{-2.9}}}
\newcommand{\loggplanetb}{\ensuremath{3.35^{+0.11}_{-0.13}}}
\newcommand{\semiMajb}{\ensuremath{0.01684^{+0.00032}_{-0.00034}}}
\newcommand{\geomAlbedob}{\ensuremath{0.61 \pm 0.17}}
\newcommand{\teqb}{\ensuremath{1833}}

\newcommand{\mplanetc}{\ensuremath{-2.80^{+6.33}_{-6.52}}}
\newcommand{\rplanetc}{\ensuremath{2.227^{+0.052}_{-0.057}}}

\newcommand{\semiMajc}{\ensuremath{0.2407^{+0.0044}_{-0.0053}}}
\newcommand{\teqc}{\ensuremath{485}}

\newcommand{\ek}{\emph{Kepler}}
\newcommand{\blender}{{\tt BLENDER}}

\shortauthors{Batalha et al.}
\shorttitle{\planetb}

\begin{document}

\bibliographystyle{apj}

\title{\emph{Kepler's} First Rocky Planet: \planetb \altaffilmark{\dagger}}

\author{
Natalie~M.~Batalha$^1$,
William~J.~Borucki$^2$,
Stephen~T.~Bryson$^2$,
Lars~A.~Buchhave$^3$,
Douglas~A.~Caldwell$^4$,
J{\o}rgen~Christensen-Dalsgaard$^{5,23}$,
David~Ciardi$^6$
Edward~W.~Dunham$^7$,
Francois~Fressin$^3$,
Thomas~N.~Gautier III$^8$, 
Ronald~L.~Gilliland$^9$,
Michael~R.~Haas$^2$,
Steve~B.~Howell$^{10}$,
Jon~M.~Jenkins$^4$,
Hans~Kjeldsen$^5$,
David~G.~Koch$^2$,
David~W.~Latham$^3$,
Jack~J.~Lissauer$^2$,
Geoffrey~W.~Marcy$^{11}$,
Jason~F.~Rowe$^2$,
Dimitar~D.~Sasselov$^3$,
Sara~Seager$^{12}$,
Jason~H.~Steffen$^{13}$,
Guillermo~Torres$^3$,
Gibor~S.~Basri$^{11}$,
Timothy~M.~Brown$^{14}$,
David~Charbonneau$^3$,
Jessie~Christiansen$^2$,
Bruce~Clarke$^4$,
William~D.~Cochran$^{15}$,
Andrea~Dupree$^3$,
Daniel~C.~Fabrycky$^3$,
Debra~Fischer$^{16}$,
Eric~B.~Ford$^{17}$,
Jonathan~Fortney$^{18}$,
Forrest~R.~Girouard$^{19}$,
Matthew~J.~Holman$^{3}$,
John~Johnson$^{20}$,
Howard~Isaacson$^{11}$,
Todd C. Klaus$^{19}$,
Pavel~Machalek$^{4}$,
Althea~V.~Moorehead$^{17}$,
Robert~C.~Morehead$^{17}$,
Darin~Ragozzine$^3$,
Peter~Tenenbaum$^4$,
Joseph~Twicken$^4$,
Samuel~Quinn$^3$
Jeffrey~VanCleve$^4$,
Lucianne~M.~Walkowicz$^{11}$,
William~F.~Welsh$^{21}$,
Edna~Devore$^{4}$,
Alan~Gould$^{22}$
}
\affil{$^1$San Jose State University, San Jose, CA 95192}
\affil{$^2$NASA Ames Research Center, Moffett Field, CA 94035}
\affil{$^3$Harvard-Smithsonian Center for Astrophysics, 60 Garden Street, Cambridge, MA 02138}
\affil{$^4$SETI Institute/NASA Ames Research Center, Moffett Field, CA 94035}
\affil{$^5$Aarhus University, DK-8000 Aarhus C, Denmark}
\affil{$^6$NASA Exoplanet Science Institute/Caltech, Pasadena, CA 91125}
\affil{$^7$Lowell Observatory, Flagstaff, AZ 86001}
\affil{$^8$Jet Propulsion Laboratory/California Institute of Technology, Pasadena, CA 91109}
\affil{$^9$Space Telescope Science Institute, Baltimore, MD 21218}
\affil{$^{10}$National Optical Astronomy Observatory, Tucson, AZ 85719}
\affil{$^{11}$University of California, Berkeley, Berkeley, CA 94720}
\affil{$^{12}$Massachusetts Institute of Technology, Cambridge, MA, 02139}
\affil{$^{13}$Fermilab Center for Particle Astrophysics, Batavia, IL 60510}
\affil{$^{14}$Las Cumbres Observatory Global Telescope, Goleta, CA 93117}
\affil{$^{15}$University of Texas, Austin, TX 78712}
\affil{$^{16}$Yale University, New Haven, CT 06510}
\affil{$^{17}$University of Florida, Gainesville, FL 32611}
\affil{$^{18}$University of California, Santa Cruz, Santa Cruz, CA 95064}
\affil{$^{19}$Orbital Sciences Corp., NASA Ames Research Center, Moffett Field, CA 94035}
\affil{$^{20}$California Institute of Technology, Pasadena, CA 91109}
\affil{$^{21}$San Diego State University, San Diego, CA 92182}
\affil{$^{22}$Lawrence Hall of Science, Berkeley, CA 94720}
\affil{$^{23}$High Altitude Observatory, National Center for Atmospheric Research, Boulder, CO 80307}
\altaffiltext{$\dagger$}{Based in part on observations obtained at the W.~M.~Keck Observatory, which is operated by the University of California and the California Institute of Technology.}

\altaffiltext{*}{To whom correspondence should be addressed.  E-mail: Natalie.Batalha@sjsu.edu}

\begin{abstract}


NASA's \ek\ Mission uses transit photometry to determine the frequency of earth-size planets in or near the habitable zone of Sun-like stars.  The mission reached a milestone toward meeting that goal: the discovery of its first rocky planet, \planetb.  Two distinct sets of transit events were detected: 1) a $\depthb$ ppm dimming lasting $\durationb$ hours with ephemeris T\,[BJD]\,$=\epochb+N*\periodb$ days and 2) a $\depthc$ ppm dimming lasting $\durationc$ hours with ephemeris T\,[BJD]\,$=\epochc+N*\periodc$ days.  Statistical tests on the photometric and pixel flux time series established the viability of the planet candidates triggering ground-based follow-up observations.  Forty precision Doppler measurements were used to confirm that the short-period transit event is due to a planetary companion. The parent star is bright enough for asteroseismic analysis. Photometry was collected at 1-minute cadence for $>4$ months from which we detected 19 distinct pulsation frequencies. Modeling the frequencies resulted in precise knowledge of the fundamental stellar properties.  \starname\ is a relatively old (\ageKASC\ Gyr) but otherwise Sun-like Main Sequence star with $\teff=\teffSME$ K, $\mstar=\mstarKASC$ \msun, and $\rstar=\rstarKASC$ \rsun.  Physical models simultaneously fit to the transit light curves and the precision Doppler measurements yielded tight constraints on the properties of \planetb\ that speak to its rocky composition: $\mpl=\mplanetb$ \mearth, $\rpl=\rplanetb$ \rearth, and $\rhopl=\rhoplanetb$ g cm$^{-3}$.  \planetb\ is the smallest transiting exoplanet discovered to date.


\end{abstract}



\section{Introduction}

NASA's \ek\ Mission, launched in March 2009, uses transit photometry to detect and characterize exoplanets with the objective of determining the frequency of earth-size planets in the habitable zone. The instrument is a wide field-of-view (115 square degrees) photometer comprised of a 0.95-meter effective aperture Schmidt telescope feeding an array of 42 CCDs that continuously and simultaneously monitors the brightness of up to 170,000 stars. A comprehensive discussion of the characteristics and on-orbit performance of the instrument and spacecraft is presented in \cite{koch10}.  The statistical properties of the stars targeted by \ek\ are described by \cite{batalha_tm}.

In January, 2010, the team announced its first 5 planet discoveries \citep{kepler4b,kepler5b,kepler6b,kepler7b,kepler8b} identified in the first 43 days of data and confirmed by radial velocity(RV) follow-up.  One of these -- Kepler-8b -- shows a clear Rossiter-McLaughlin velocity variation which allowed for the measurement of the spin-orbit alignment of the system \citep{kepler8b}.  The ``first five'' are all short-period giant planets, the smallest being comparable in size to Neptune. Collectively, they are similar to the sample of transiting exoplanets that have been identified to date\footnote{\url{http://exoplanet.eu/}}, the ranks of which currently hover around 100.  The median mass of the sample is 0.99 \mjup\ with a 10$^{th}$ and 90$^{th}$ percentile of 0.24 \mjup\ and 4.1 \mjup.  The median radius is 1.18 \rjup\ with a 10$^{th}$ and 90$^{th}$ percentile of 0.81 \rjup\ and 1.5 \rjup.  The known transiting planets are, statistically speaking, Jovian-like in both mass and size and have short orbital periods with a median value of 3.5 days and a 10$^{th}$ and 90$^{th}$ percentile of 1.5 days and 8.0 days, respectively.  

In June 2010, \ek\ released a catalog of 306 stars with planet-like transit signatures \citep{Borucki:10}.  Even if the majority turn out to be false positives, the number of transiting planets could plausibly more than double from this pool of candidates.   Soon, we will leave the realm of small-sample statistics and be able to say something meaningful about not only the mass and size distribution but also the dynamical and compositional nature of exoplanets.  Information about composition will fall from those systems for which we can derive not only a radius, but also a mass.  Dynamical information will fall from multiple-transiting planet systems.  Five such candidate systems were included in the catalog of \citet{Borucki:10} and described in more detail by \citet{multis}.  The discovery of the planets orbiting Kepler-9 \citep{kepler9bc} marked the first confirmation of a multiple-transiting planet system.  Kepler-9 is a G-type star with two Saturn-mass transiting planets in a near $2:1$ orbital resonance.  The system is also the first to show transit timing variations.  Dynamical models of these variations afford us the means of determining planetary mass without the need for RV follow-up.  

After removing the transit signals of Kepler-9b and Kepler-9c, a third transit signature was identified in the light curve revealing an additional candidate with a 1.6-day period.  The planetary interpretation of Kepler-9d was validated \citep{Torres:10} without detection of a Doppler signal.  Rather, given the star and transit properties, a matrix of possible false-positive scenarios was constructed.  After eliminating all scenarios which are not consistent with the observables, a false-alarm probability was computed that speaks to the likelihood that Kepler-9d is consistent with the planet interpretation.  In this manner, Kepler-9d was validated with high confidence as a super-earth-size planet with radius $1.64^{+0.19}_{-0.14}$ \rearth.  

Here, we attempt to define ``super-earth'' from a radius perspective by noting that the 10 \mearth\ upper limit proposed by \citet{valencia:06} corresponds roughly to 2 \rearth\ for a planet with no water and low Fe/Si ratio \citep{zeng}.  Practically speaking, the term super-earth has been loosely used to refer to all planets larger than earth and smaller than Neptune.  However, this is a broad domain that captures not only rocky, dry planets that happen to be larger than Earth, but also ocean planets and mini-Neptunes.  These finer distinctions will only be possible with measurements of planetary properties better than $5\%$ in radius and $10\%$ in mass \citep{valencia:06,valencia:07,fortney,seager:07,grasset}. Pending a mass determination, there is no information as of yet with regards to the composition of Kepler-9d.

The discovery of a short-period, super-earth-size planet is not surprising.  Indeed, one of the most interesting aspects of the \ek\ candidate sample reported by \citet{Borucki:10} is the fact that the median of the radius distribution is strikingly different than that of the known transiting exoplanets.  The \ek\ candidates have a median radius of 0.30 \rjup\ -- smaller than that of Neptune (0.34 \rjup) -- and candidates as small as 1.5 \rearth.  Should this distribution survive the process of false positive elimination, we will see substantial numbers of short-period super-earths.  There is precedence already for transiting exoplanet discoveries in the super-earth domain.  CoRoT-7b \citep{leger,queloz,bruntt,pont} is an example: a $1.58\pm0.10$ \rearth, $4.8\pm0.8$ \mearth\ planet orbiting a K-type star.  The case for CoRoT-7b is complicated by activity-induced RV jitter.  Independent analysis of the Doppler measurements by \citet{pont} reduces the significance of the detection somewhat (yielding $2.3\pm1.8$ \mearth) and puts the 95\% confidence interval between 0 and 5 \mearth\ -- still, however, within the super-earth domain.  And while the \citet{queloz} mass and radius point to a rocky composition, the lower mass of \citet{pont} marginally favors a water/ice composition.

GJ 1214b \citep{charbonneau} is another example of a transiting super-earth (as defined from a mass perspective) at 6.55 \mearth.  Orbiting an M-type star with a period of 1.58 days, GJ1214b has a radius of 2.68 \rearth.  Its density (1.87 g cm$^{-3}$), consequently, is closer to that of water than that of the Earth.  The interior structure of GJ 1214b has been modeled as an H/He/H$_2$O planet with a rocky core \citep{nettelmann}.

Here, we report on the discovery of a super-earth-size exoplanet orbiting the G-type Main Sequence star, \starname\ (\kicid).  At Kp$=\kepmag$, the star is bright enough for asteroseismic analysis of its fundamental stellar properties using the high precision \ek\ photometry.  The stellar properties are known to an accuracy that allows us to put \planetb\ sitting squarely in the rocky domain of the mass-radius diagram.  The light curve shows two distinct sets of transit events: one at $\periodb$ days, referred to as \koib\, and the other at $\periodc$ days, referred to as \koic\ where ``KOI'' denotes a \ek\ Object of Interest.  

The \starname\ data acquisition, photometry and transit detection are described in Section~\ref{sec:photometry}.  The statistical tests performed on the \ek\ photometry to rule out false positives are described in Section~\ref{sec:dv}, and the subsequent ground-based observations, including precision Doppler measurements, leading to the confirmation of \koib\ are described in Section~\ref{sec:fop}.  Throughout the first half of this paper, we refer to each event as \koib\ and \koic.  However, in the latter half, we begin to discuss \koib\ in the context of a confirmed planet and call it out accordingly as \planetb.  The \koic\ signal, at this time, has not been confirmed by RV.  Many of the false-positive scenarios have been investigated via \blender\ analysis \citep{Torres:10} as described in Section~\ref{sec:blender}.  \koic\ will require analysis beyond the scope of this paper for validation at an acceptable confidence level, since eliminating possible astrophysical false positives is more difficult given the possibility of an eccentric orbit.  The larger orbital separation of \koic\ precludes us from assuming that tidal effects will have circularized the orbit.  While both the transit duration and RV observations are consistent with a circular orbit, the uncertainties are large enough that a much more comprehensive \blender\ analysis will be required to validate the outer planet candidate.  Thus, we refer to this candidate throughout the paper as \koic.

From spectroscopy to asteroseismology, the analyses yielding fundamental stellar properties are discussed in Section~\ref{sec:star}.  Section~\ref{sec:planet} contains a description of the light curve plus radial velocity modeling that yield the planet properties of \planetb\ (and \koic\ under the planet interpretation). In the case of \koic\, the absence of an RV signal implies an upper mass limit.  We look for small, systematic deviations in the transit arrival times that could indicate the dynamical interaction of multiple planets orbiting \starname.  The implications of a null detection are discussed in Section~\ref{sec:ttv}.  Finally, the properties of \planetb\ are discussed in the context of the theoretical models that speak to the planet's composition.  Its placement in a mass-radius diagram suggests a dry, rocky composition.  Moreover, the high density ($\rhoplanetb$ g cm$^{-3}$) indicates a large iron mass fraction as discussed in Section~\ref{sec:discussion}.  

\emph{Kepler's} primary objective is to determine the frequency of earth-size planets in the habitable zone.  The number of planet candidates identified in $< 1$ year of photometry is fast approaching the thousands.  Time on the large telescopes required for precision Doppler measurements is not only costly, it is insufficient for confirming earth-size planets in the HZ of sun-like stars. Instruments are not yet capable of yielding the cm s$^{-1}$ precision required for confirmation of \emph{Kepler's} most interesting candidates.  For the near-term future, the team will focus its efforts on quantifying the false-positive rate well enough that it might be applied to the collective sample of planet candidates.  The observations of \starname\ are part of that effort.  The discovery of \planetb\ marks an important milestone for the team:  \emph{Kepler's} first rocky planet and the smallest transiting exoplanet discovered to date.

\section{\ek\ Photometry}
\label{sec:photometry}

The discovery of the planet orbiting \starname \ begins with the high-precision \ek\ photometry.  Indeed not one but two periodic transit events were identified in the light curve, producing pipeline statistics that initiated the cascade of verification and follow-up efforts leading to confirmation and characterization of \planetb.
 
\subsection{Data Acquisition}
\label{sec:data}

The \ek\ photometer is a 0.95-m effective aperture, wide field of view Schmidt camera in an Earth-trailing orbit.  It is designed to yield 20 parts per million relative time series precision in 6.5-hours for a 12th magnitude G2 star.  The focal plane is comprised of 42 1024x2200 pixel science CCDs arranged together in 21 roughly square modules covering 115 square degrees of sky (3.98 arcsec per pixel).  Each pair of CCDs forming a module shares a common sapphire field-flattener lens.  The coatings deposited onto the field-flattener lenses (and, to a lesser degree, the optics and quantum efficiency of the CCDs) define the effective bandpass of the otherwise filter-less photometer, yielding a mean transmission of 52.6\% between 423 and 897 nm (defining the 5\% transmission points).  Each CCD requires its own transmission function for meaningful interpretation of color-dependent behaviors that affect planet characterization (e.g. limb darkening). A description of the instrument is given in \citet{vancleve} and \citet{argabright08} while an overview of its in-flight performance is presented in \citet{caldwellSPIE} and \citet{jenkins:10}.
  
Each CCD is electronically divided into two 1024x1100 output units defining a total of 84 readout channels.  While the photometer points at a single field continuously throughout its heliocentric orbit, it rotates about the optical axis once every $\sim$3 months (hereafter referred to as a Quarter) in order to keep the solar panels facing the Sun. Consequently, every star spends each quarter of the year on a different channel.  \starname \ (RA$=\kicra$, DEC$=\kicdec$) falls on channels 36 (module 11, output 4), 80 (module 23, output 4), 52 (module 15, output 4), and 8 (module 3, output 4) in quarters 0/1 (spring), 2 (summer), 3 (fall), and 4 (winter), respectively. 

CCDs are read out every $\sim6.5$ seconds ($6.01982$ second integration and $0.51895$ second read time), and every 270 readouts are co-added onboard to form $1765.5$ second ($\sim 29.4$ minute) integrations (long-cadence). Data for up to $170,000$ stars are recorded at long-cadence (LC) while data for up to 512 stars are also co-added to a $58.85$ second ($\sim 1$ minute) integration (9 readouts) termed short-cadence (SC) as described in \cite{jenkins:10,gilliland:10}.  The LC photometry of \starname \ used in the analyses reported here was acquired between 02 May 2009 and 09 January 2010 -- quarters 0/1 (spring) up through the first month of quarter 4 (winter). In early January, the module containing channel 8 experienced a hardware failure that was not recoverable and observations of \starname \ were cut short until the subsequent spacecraft roll.  Over 11,000 long-cadence observations are used in this analysis.  Short cadence data were also collected between 21 July - 19 August 2009 (one month of quarter 2) and between 18 September 2009 through 09 January 2010 (quarter 3 and the first weeks of quarter 4).  The short-cadence data were vital in determining the fundamental stellar parameters from an asteroseismic analysis (p-mode detection) described in Section~\ref{sec:astero}.  Approximately 200,000 SC observations were collected in this time period. Both LC and SC data are used in the light curve modeling carried out to characterize the planets. Observations of \starname \ continue in the three quarters each year when the target is not on the failed module.

\subsection{Light Curves}
\label{sec:lightCurves}

\begin{figure}
\begin{center}
\includegraphics[width=175mm]{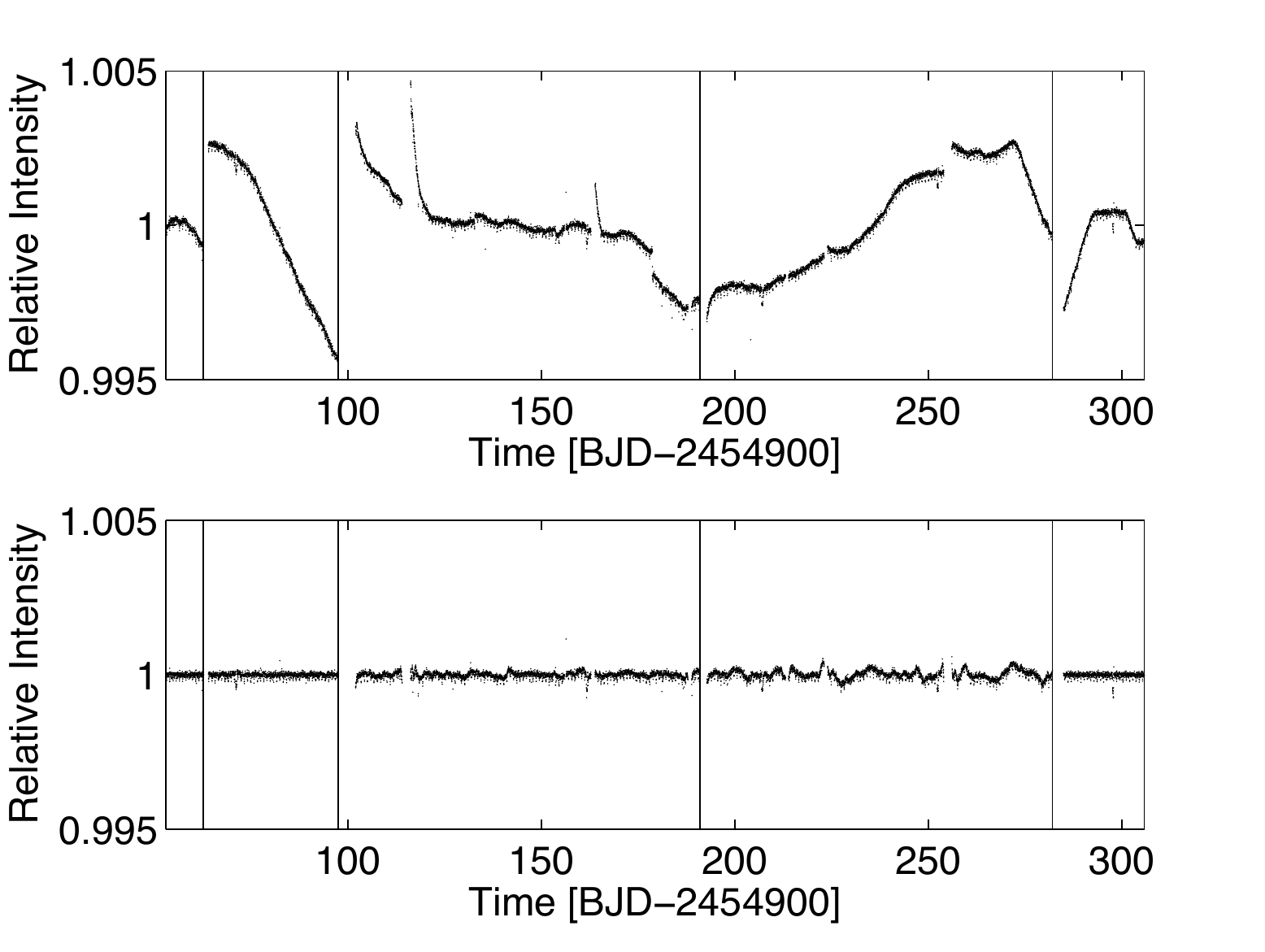}
\end{center}
\caption{Raw (upper) and corrected (lower) flux time series produced by the \ek\ photometry pipeline (PA and PDC fluxes, respectively).  Vertical lines denote the boundaries between quarters. Intra-quarter flux values have been scaled by their median flux for display purpose only (to mitigate the discontinuities between quarters).}
\label{fig:rawFlux}
\end{figure}

Raw flux light curves are extracted by performing an unweighted sum of calibrated pixels that have been subjected to cosmic ray removal and background subtraction \citep{socpipeline}. The pixels used in the sum are those defining the optimal aperture -- the set of pixels that optimizes the total signal-to-noise ratio (SNR). The optimal aperture is dependent on the local Pixel Response Function\footnote{The Pixel Response Function (PRF) is a super-resolution representation of the distribution of starlight over the CCD pixels.  It includes not only the effects of the instrumental optics, but also intra-pixel sensitivity and pointing jitter.} (PRF), measured on-orbit during the commissioning period \citep{Bryson:10}, the distribution of stellar flux on the sky near the target (crowding), and differential velocity aberration. A complete discussion of \emph{Kepler's} aperture photometry pipeline (PA) is given in \cite{pa}. Systematic errors, outliers and intra-quarter discontinuities are removed by cotrending against ancillary data products in the Pre-Search Data Conditioning (PDC) pipeline module as described in \cite{pdc}. Figure~\ref{fig:rawFlux} shows the raw (upper panel) and corrected (lower panel) flux time series for \starname.  Vertical lines denote the boundaries between quarters.  Intra-quarter fluxes were normalized by their median flux in order to reduce the magnitude of the flux discontinuities between quarters.  The largest systematic errors are the long-term drifts due to image motion (differential velocity aberration) and the thermal transients after safe mode events (e.g. that near day 115). After filtering out transit events, the measured relative standard deviation of the PDC-corrected, long-cadence light curve is $62$ ppm per (LC) cadence.  An expected instrument $+$ photon noise is computed for each flux in the time series.  The mean of the per (29.4-minute) cadence noise estimates reported by the pipeline is 36 ppm. Both raw (PA) and corrected (PDC) light curves are available at the Multi-Mission Archive at Space Telescope Science Institute MAST\footnote{\url{http://archive.mast.edu/kepler}}. 

With regards to the time stamps associated with each photometric flux, we note that \emph{Kepler's} fundamental coordinate system is UTC for all time tags.  Spacecraft times are converted to barycentric-corrected Julian Dates at the mid-time of each cadence.

\subsection{Transiting Planet Search}
\label{sec:tps}

The \planetb \ transits were identified by the Transiting Planet Search (TPS) pipeline module that searches through each systematic error-corrected flux time series for periodic sequences of negative pulses corresponding to transit signatures. The approach is a wavelet-based, adaptive matched filter that characterizes the power spectral density of the background process (i.e. anything that is not related to 
the transit signal itself, such as stellar variability, instrumental signatures, differential velocity aberration-induced photometric drift
, etc) yielding the observed light curve and uses this time-variable power spectral density estimate to realize a pre-whitening filter to apply to the light curve \citep{jenkinsSPIE}. TPS then convolves a transit waveform, whitened by the same pre-whitening filter as the data, with the whitened data to obtain a time series of single event statistics. These represent the likelihood that a transit of that duration is present at each time step. The single event statistics are then combined into multiple event statistics by folding them at trial orbital periods ranging from 0.5 days to as long as one quarter ($\sim$90 days). The step sizes in period and epoch are chosen to control the minimum correlation coefficient between neighboring transit models used in the search so as to maintain a high sensitivity to transit sequences in the data.  

\koib\ was identified by TPS in each quarter of data with a multiple event statistic $>15\sigma$. The long-period transits of \koic\ were identified by manual inspection due to the fact that the current version of TPS does not operate on more than one quarter of data at a time making detection of events with periods beyond $\sim 30$ days necessarily incomplete.  Multi-quarter functionality is slated for the next software release in early 2011.  The transit depth, duration, period, and epoch are derived from physical modeling (see Section~\ref{sec:planet}) using all of the available data.  \koib\ is characterized as a \depthb\ ppm dimming lasting \durationb\ hours with transit ephemeris of T\,[BJD]\,$=\,\epochb\,+\,N*\periodb$ days. \koic\ is characterized as a \depthc\ ppm dimming lasting \durationc\ hours and an ephemeris T\,[BJD]\,$=\,\epochc\,+\,N*\periodc$ days.

\section{Data Validation}
\label{sec:dv}

Astrophysical signals mimicking planet transits are routinely picked up by the pipeline.  The large majority of such false positives can be identified via statistical tests performed on the \ek\ data itself -- tests that are collectively referred to as {\it Data Validation}. Data Validation for \emph{Kepler's} first planet discoveries (Kepler-4b, 5b, 6b, 7b, and 8b) is described in \cite{batalha_fp}. These are gradually being replaced by pipeline software products such as those described in \cite{dv}. Only targets passing each of these statistical tests are passed on to the Follow-up Observation Program team for further vetting, confirmation, and characterization.  It is at this stage that stars are assigned a ``\ek\ Object of Interest'' (KOI) number.  \starname \ was referred to as KOI-72 throughout the vetting stages.  More specifically, the short-period event was referred to as \koib \ while the long-period event was referred to as \koic.  We will use these identifiers in the subsequent discussions of the analyses that led to confirmation. Here we describe the Data Validation metrics -- statistics which, taken alone, support the planet interpretation for both \koib \ and \koic.

\subsection{Binarity Tests}
\label{sec:binarity}

For each event, the even-numbered transits and odd-numbered transits are modeled independently using the techniques described in Section~\ref{sec:planet}.  The depth of the phase-folded even-numbered transits is compared to that of the odd-numbered transits as described in \cite{batalha_fp}.  A statistically significant difference in the transit depths is an indication of a diluted or grazing eclipsing binary system.  Neither of the transit events detected in the light curve of \starname \ shows odd-even depth differences outside of $2\sigma$ where $\sigma$ refers to the uncertainty in the transit depths reported in Section~\ref{sec:tps} (6 and 9 ppm for \koib\ and \koic\, respectively). 

The modeling allows for the presence of a secondary eclipse (or occultation event) near phase=0.5 and reports the significance of such a signal. While its presence does not rule out the planetary interpretation, it acts as a flag for further investigation.  More specifically, the flux decrease is translated into a surface temperature assuming a thermally radiating disk, and this temperature is compared to the equilibrium temperature of a low albedo (0.1) planet at the modeled distance from the parent star.  There is a marginal detection ($\sim 2\sigma$) of a secondary eclipse associated with \koib. At just 6 ppm, the flux change is not severe enough to rule out the planetary interpretation (i.e. that it is due to an occultation of a planetary companion).  There is no detection of a secondary eclipse associated with \koic.  The binarity tests are consistent with the planet interpretation for both transit events in the light curve of \starname.

\subsection{Photocenter Tests}
\label{sec:centroid}

To check for false positives due to background eclipsing binaries, we study the behavior
of flux centroids -- the center-of-light distribution in the photometric aperture -- and how it behaves as a function of time, especially comparing images taken during transits with those taken outside of transit.  We rely primarily on flux-weighted centroids and modeling the expected behavior based on the local distribution of point sources from the \ek\ Input Catalog and supplemented by high spatial resolution imaging (Section~\ref{sec:imaging}).  Neither Speckle nor AO imaging reveal any point sources in the photometric aperture that were not already listed in the \ek\ Input Catalog. 

The study of centroid behavior is complicated by the fact that the \starname\ image is saturated in all quarters.
In Quarter 2, the star is on the edge of saturation, but it is apparent that there is some spilling of charge even in this case.
Saturation behavior on the \ek\ focal plane is known to be conservative (e.g. all charge is captured)
but has strong pixel-to-pixel variation in details such as saturation level and the fraction of flux that
spills up and down the CCD columns.  

The high-accuracy technique of fitting the Pixel Response Function (PRF) to the difference image
formed by subtracting the in-transit image from the out-of-transit image is inappropriate for saturated
targets, because the saturated pixels do not represent the PRF of the star, and such mismatches between
the pixel data and the PRF cause significant position biases in the PRF fit.  As an alternative, we use a modified
PRF fit technique, described below.

We compute flux-weighted centroids by
creating out-of-transit and in-transit images from detrended, folded
pixel time series.  Separate images are created for \koib \ and
\koic.  For each pixel time series, the de-trending
operation has three steps: 1) removal of a median filtered
time series with a window size equal to the larger of 48 cadences or three times
the transit duration; 2) removal of a robust low-order polynomial fit; and 3)
the application of a Savitzky-Golay filtered time series of order 3
with a width of 10 cadences (5 hours).  The Savitzky-Golay filter is not applied
within 2 cadences of a transit event, so the transits are preserved.
The resulting pixel time series are folded by the transit period.
Each pixel in the out-of-transit image is the average of 30 points
taken from the folded time series outside the transit, 15 points on
each side of the transit event.  Each pixel in the in-transit image is
the average of as many points in the transit as possible: three for
\koib \ and eleven for \koic. In the former case, 
though there are fewer points per transit, there are significantly more transits to draw from
(several hundred for \koib \ compared to just six for \koic).
Consequently, \koib \ yields smaller uncertainties.

A flux-weighted centroid is computed for the out-of-transit image and
the in-transit-image using all of the pixels in the 
mask\footnote{Each star observed by \ek\ has a predefined mask that defines the pixel 
set that is downloaded from the
spacecraft. The mask changes from quarter to quarter since the star falls on a different 
CCD channel upon quarterly spacecraft rotation.} of \starname. This
produces row and column centroid offsets $\Delta R$ and $\Delta C$,
and the centroid offset distance $D = \sqrt{\Delta R^2 + \Delta C^2}$.

Uncertainties of these centroids are estimated via Monte Carlo
simulation, where a noise realization is injected into 48-cadence
smoothed versions of the pixel time series for each trial. A total of
2000 trials are performed each for \koib\ and
\koic.  The in- and out-of-transit images are formed using the
same de-trending, folding and averaging as the flight data.  The
measured uncertainties are in the range of a few times $10^{-5}$
pixels.

Table~\ref{tab:measured_centroids_01} shows the resulting measurements of
the \koib \ centroids from quarter 1, 2, 3, and 4 pixel data, along with the
Monte-Carlo-based $1\sigma$ uncertainties.  The centroids are
converted into centroid offsets and offset distance with propagated
uncertainties.  We see that while in quarter 1 there is a $> 4 \sigma$ observed
offset, in quarters 2 and 3 the observed offset is less than $1 \sigma$.  Quarter 1
had significant thermal and pointing systematics, while quarter 2 had significant
pointing drift.  These systematics were significantly reduced
by quarter 3.

Table~\ref{tab:measured_centroids_02} shows the measured centroid shifts
and uncertainties for
\koic \ in quarters 1 through 4.  The \koic \ transit uncertainties in quarter 3 are smaller
because only one transit was observed in quarters 1, 2, and 4
while two transits were observed in quarter 3.  Generally, the centroid shifts for \koic \
are larger than \koib.  This larger centroid shift may be indicative of
the \koic \ transit event being due to a background eclipsing binary, or the larger shift
may be due to the deeper transits of \koic \ (larger flux changes induce larger centroid shifts) combined with the
complex behavior of pixels at or near saturation.   To study this question
PRF-fit centroids that ignore saturated pixels are computed for
the \koic \ in- and out-of-transit images, and the resulting centroid shifts are computed.
These results are shown in  Table~\ref{tab:prf_centroids_02}.  This technique
indicates no statistically detectable centroid motion in quarter 3.  The
uncertainties were computed via Monte-Carlo methods similar to those used to
compute the flux-weighted centroid uncertainties.

While both the flux-weighted centroid and PRF fitting methods indicate statistically
significant centroid motion on \koic \ in quarters 1, 2, and 4, we point out that
the directions of the centroid offsets are inconsistent from quarter to quarter.
This indicates that these centroid offsets are not likely to be due to a
background object in the sky.

The observed centroids are compared with the modeled centroids computed using 
point sources catalogued in the \ek\ Input Catalog \citep[KIC;][]{kicLatham} 
out to 15 pixels beyond the mask. As described in Sections~\ref{sec:speckle} and \ref{sec:ao}, 
no additional point sources were identified in Speckle or AO imaging. 
To generate the modeled out-of-transit
image, the measured PRF is placed at each star's location on the focal
plane, scaled by that star's flux.  This provides the contribution of
each star to the flux in the mask.  For each star $s_i$
in the mask, the depth $d_{s_i}$ of a transit is computed that reproduces the
observed depth in the individual pixels.  An in-transit
image for each $s_i$ is similarly but with $s_i$'s flux suppressed by $1-
d_{s_i}$.  These model images are subject to errors in the PRF
\citep[][]{Bryson:10}, so they will not exactly match the sky.  Further, a very simple
saturation model is applied which spills saturation symmetrically up and down the
column.

The modeled centroids (using the quarter 3 mask definition) are presented in Table~\ref{tab:modeled_offsets}.  
For both \koib\ and \koic\, the offset expected for a transit associated with \starname\ (and not a nearby star)
is smaller than the $1 \sigma$ uncertainties in $D$ computed from the flux-weghted centroid
measurements.  Modeled transits on other stars in the mask predict centroid shift in excess of $10 \sigma$,
which would be readily observed in our flux-weighted centroids.  We can therefore rule out
all known stars in the mask as responsible for the transit signal on both \koib \ and \koic.

There is a radius beyond which any star capable of producing the observed transit signal would also induce a centroid shift large enough to be detected in the \ek\ data when comparing in- and out-of-transit images.  We estimate this confusion radius by scaling the $3 \sigma$ centroid offset uncertainty by the observed transit depth as described in Section 4.1.3 of \cite{dv}.  For this, we use the quarter 3 uncertainty in $D$.  This radius of confusion for \koib \ is $1.17$ arcsec, and the radius of confusion for \koic \ is $0.60$ arcsec.  The volume subtended by this area on the sky can be used in a \blender\ analysis (Section~\ref{sec:blender}) to assess the probability of encountering an eclipsing binary (capable of producing the transit signal) in that volume of the Galaxy. However, in the case of \starname\, the high spatial resolution imaging described in Sections~\ref{sec:speckle} and \ref{sec:ao} provides tighter constraints on the background star population.

\section{Follow-Up Observations}
\label{sec:fop}

Each of the periodic transit signals identified in the light curve of \starname\ passes all of the Data Validation tests that might indicate the possibility of a false-positive as described in Section~\ref{sec:dv}.  The star was passed to the follow-up observing team on July 21, 2009 after identification and scrutiny of the short-period transit event (\koib).  This initiated a series of ground-based observations that began with reconnaissance spectroscopy to confirm the stellar parameters in the \ek\ Input Catalog and identify any obvious eclipsing binary signatures (Section~\ref{sec:recon}), continued with high spatial resolution imaging to identify nearby stars in the photometric aperture (Section~\ref{sec:imaging}), and ended with high-resolution, high SNR echelle spectroscopy with and without an iodine cell to compute stellar parameters, probe magnetic activity, measure line bisectors, and make precision Doppler measurements.  The follow-up observations do not rule out the planetary interpretation for either of the transit signatures.  However, they only allow for the confirmation and characterization of the short-period candidate, \koib\ as discussed in Section~\ref{sec:rv}.

\subsection{Reconnaissance Spectroscopy}
\label{sec:recon}

Spectroscopic observations on medium-class telescopes are acquired
before requesting precision Doppler measurements.  These
``reconnaissance'' spectra are used to improve upon the
photometrically-derived stellar classification from the \ek\ Input
Catalog ($\teff = \kicteff$ K, $\logg = \kiclogg$, $\rstar =
\kicradius \rsun$), identify double-lined spectroscopic binaries, and
search for indications of radial-velocity variations larger than $\sim
1$ km s$^{-1}$ that might suggest a stellar companion. The objective is
continued false-positive elimination.

Two reconnaissance spectra were obtained with the Hamilton echelle on
the Shane 3-m telescope at the Lick Observatory.  High SNR observations
were acquired on two successive nights in August 2009, at heliocentric
Julian dates 2455046.771 and 2455047.758, corresponding to phases
0.146 and 0.3251 for \koib, and to phases 0.658 and 0.680 for \koic.
The spectral order covering about 7.0 nm centered on the Mg b lines was
correlated against a library of synthetic spectra calculated by John
Laird using a line list prepared by Jon Morse.  The stellar parameters
for the template spectrum that yielded the highest value for the peak
correlation coefficient were $\teff = \teffReconOrig$ K, $\logg = \loggReconOrig$,
and $\vsini = \vsiniReconOrig$ km s$^{-1}$ for an assumed solar
metallicity.  The errors quoted for \teff\ and \logg\ are half the spacing of
the library grid.  The two Lick exposures were unusually high SNR compared to typical
reconnaissance spectra, with a peak value for the correlation
coefficient of 0.98.  Consequently, we were able to estimate the metallicity and
to interpolate to finer values of the temperature and gravity,
obtaining $\teff = \teffRecon$ K, $\logg = \loggRecon$, and $\feh =
\fehRecon$.  The correlation functions for the two Lick observations
showed no evidence of a composite spectrum, and the two velocities
agreed within 0.1 \kms.  Thus there was no suggestion of a stellar
companion responsible for either system of transit events, and the
reconnaissance spectroscopy confirms that \starname\ is a sun-like,
slowly rotating Main Sequence star, and supports the planetary
interpretation for the transit events. Consequently, the star was
scheduled for precision Doppler measurements (see
Section~\ref{sec:rv}).

\subsection{High Spatial-Resolution Imaging}
\label{sec:imaging}

The more complete our knowledge of stellar flux sources in the photometric aperture, the better we are able to assess the likelihood of a blend scenario in the interpretation of the transit event.  Much of this knowledge comes from the \ek\ Input Catalog (KIC) which federates point sources from the USNO-B catalog, the 2MASS catalog, and our own pre-launch Stellar Classification Program.  Identification of point sources within a 1.5" radius requires additional imaging. 

\subsubsection{Speckle Imaging}
\label{sec:speckle}

\begin{figure}
\begin{center}
\includegraphics[height=150mm]{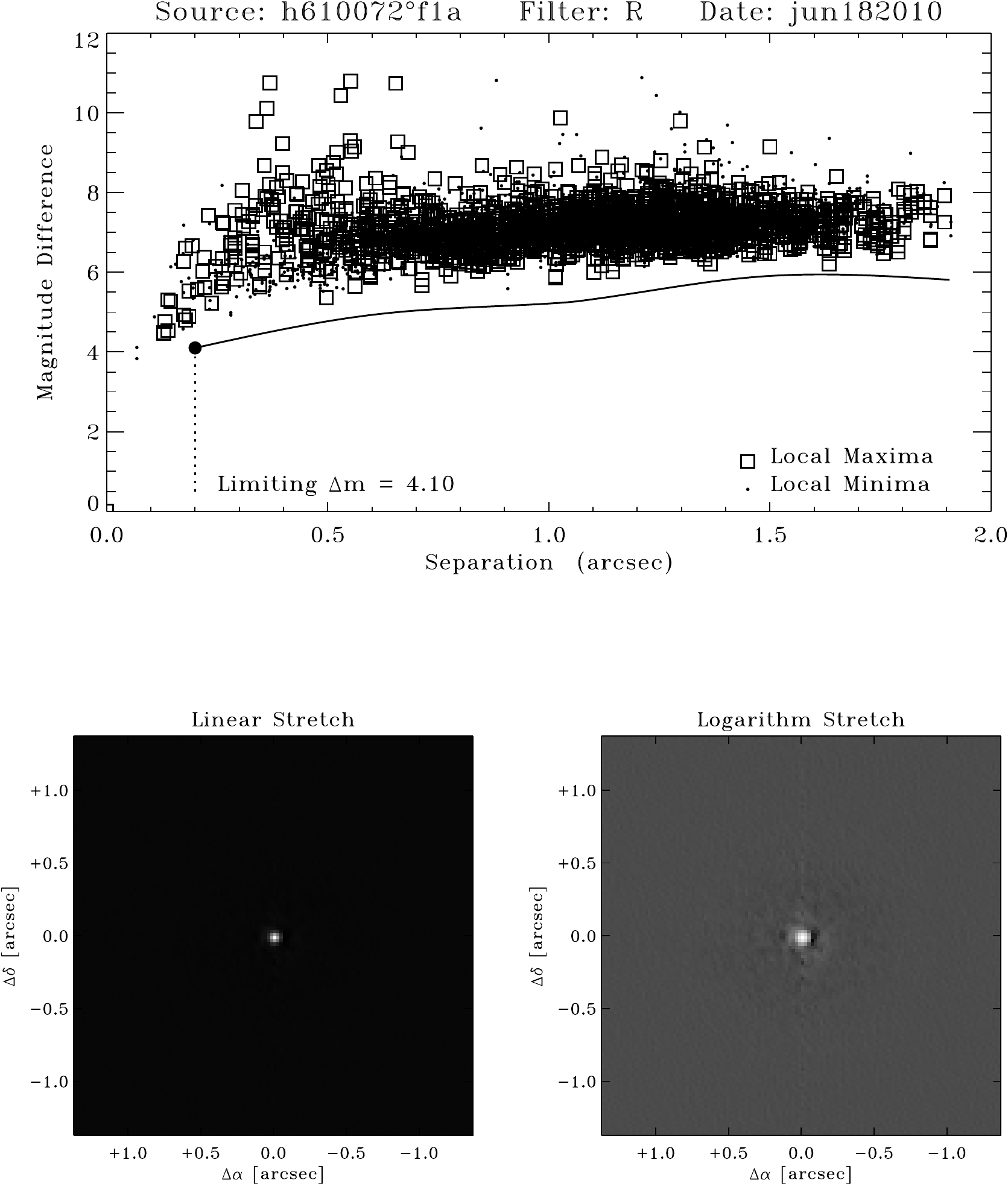}
\end{center}
\caption{Speckle reconstructed $R$ band image of \starname.  No other source is observed to within 1.8 arcseconds of the target to a depth of $\sim$6 magnitudes. The cross pattern is an artifact of the reconstruction process.  North/East are up/left.}
\label{fig:speckle}
\end{figure}

Speckle imaging of \starname\ was obtained on the night of 18 June 2010 UT using the two-color speckle camera at the WIYN 3.5-m telescope located on Kitt Peak. The speckle camera simultaneously obtained 2000 30 msec EMCCD images in two filters: $V$ (5620/400\AA) and $R$ (6920/400\AA). These data were reduced and processed to produce a final reconstructed speckle image for each filter. Figure~\ref{fig:speckle} shows the reconstructed R band image.  North is up and East is to the left in the image and the ``cross'' pattern seen in the image is an artifact of the reconstruction process.  The details of the two-color EMCCD speckle camera are presented in \citet{howell10}.

For the speckle data, we determine if a companion star exists within the approximately $2.5 \times 2.5$ arcsec box centered on the target and robustly estimate the background limit we reach in each summed, reconstructed speckle image. The two-color system allows us to believe single fringe detection (finding and modeling identical fringes in both filters) if they exist and rule out companions between 0.05 arcsec and 1.5 arcsec from \starname.  The speckle image was obtained with the WIYN telescope native seeing near 0.7 arcsec, and we find no \starname\ companion star within the speckle image separation detection limits to a magnitude limit of 6 mag in R and $4.5$ in mag in V below the brightness of \starname.

\subsubsection{AO Imaging}
\label{sec:ao}

\begin{figure}
\hspace{100pt}
\includegraphics[height=150mm]{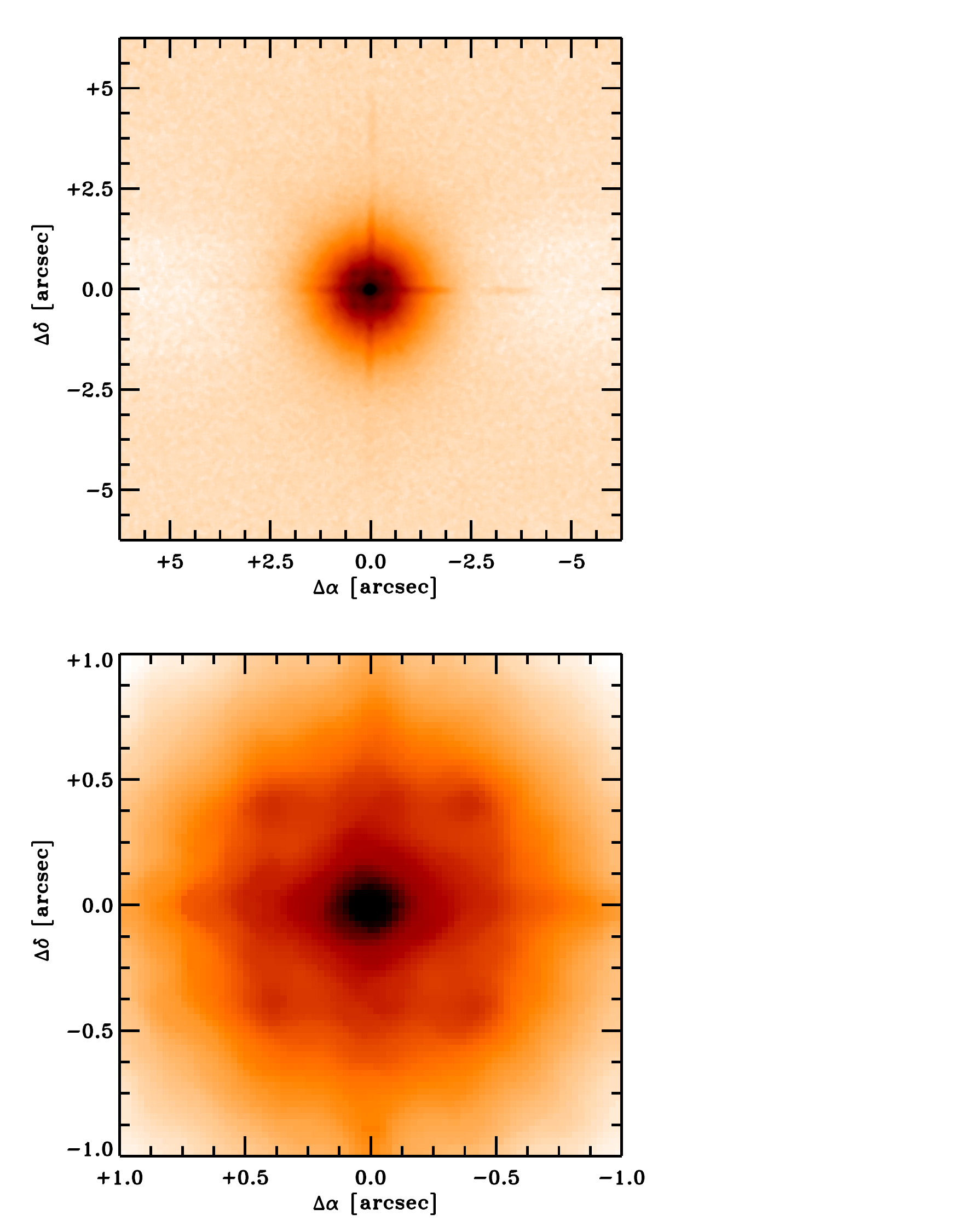}
\caption{$J$-band Palomar adaptive optics image of \koi.  The top image displays a
$12.5^{\prime\prime} \times 12.5^{\prime\prime}$ field of view centered on the
primary target.  The bottom image displays a  $2^{\prime\prime} \times
2^{\prime\prime}$ field of view centered on the primary target. The four-point
pattern surrounding the central point spread function core is part of the
adaptive optics point spread function.}
\label{fig:paloAO}
\end{figure}

Near-infrared adaptive optics imaging of \koi \ was obtained on the night
of 08 September 2009 UT with the Palomar Hale $200''$ telescope and the PHARO
near-infrared camera \citep{hayward2001} behind the Palomar adaptive optics
system \citep{troy2000}.  PHARO, a $1024\times1024$ HgCdTe infrared array,
was utilized in 25.1 mas/pixel mode yielding a field of view of
$25$ arcsec.  Observations were performed using a $J$ filter ($\lambda_0 =
1.25\mu$m).  The data were collected in a standard 5-point quincunx dither
pattern (e.g. dice pattern for number five) of $5$ arcsec steps interlaced with an off-source 
($60$ arcsec East) sky dither pattern. Data were taken at two separate times within the
same night -- 150 frames using 1.4-second integration times and 150 frames using 
2.8-second integration times --  for a total on-source integration time of 10 minutes.  The
individual frames were reduced with a custom set of IDL routines written
for the PHARO camera and were combined into a single final image.  The
adaptive optics system guided on the primary target itself and produced
measured Strehl ratios of 0.15 at $J$ with a central core width of $FWHM =
0.075$ arcsec.  The final coadded image at $J$ is shown in
Figure~\ref{fig:paloAO}.

No additional sources were detected at $J$ within $6.25$ arcsec of the
primary target.  Source detection completeness was accomplished by randomly
inserting fake sources of various magnitudes in steps of 0.5 mag and at
varying distances in steps of 1.0 FWHM from the primary target.
Identification of sources was performed both automatically with the IDL
version of DAOPhot and by eye. Magnitude detection limits were set when a
source was not detected by the automated FIND routine or 
by eye.  Within a distance of $1-2$ FWHM, the automated finding routine
often fails even though the eye can discern two sources. Beyond that
distance the two methods agreed well.  A summary of the detection
efficiency as a function of distance from the primary star is given in
Table~\ref{tab:paloAO}.

\subsection{Precise Doppler Measurements of \starname}
\label{sec:rv}

We obtained 40 high resolution spectra of \starname\ between 2009 Aug 31
and 2010 Aug 06 using the HIRES spectrometer on the Keck~I 10-m
telescope \citep{vogt94}.  We used the same configuration of HIRES that is
normally used for precise Doppler work of nearby FGK stars
\citep{Marcy08} which yields a Doppler precision of 1.0-1.5 \ms\ 
depending on spectral type and rotational \vsini.  The HIRES fiber-feed was not used for these observations. The standard iodine cell was placed in the telescope beam to superimpose iodine lines
directly on the stellar spectrum.  As the iodine lines and stellar
lines are carried by exactly the same photons hitting the same optics, both sets of lines share precisely the same instrumental profile and wavelength scale.   The iodine lines represent exactly the same spectrometer optics as the stellar spectrum, with no difference. We fit the composite spectrum of iodine and
 stellar lines simultaneously in each 100 pixel segment
of spectrum, yielding a Doppler shift that automatically includes the
instantaneous wavelength scale and instrumental profile.  This
dual-fitting limits the long-term and short-term systematic Doppler
errors at $\sim$1.0 \ms.

Most observations were made with the ``C2 decker'' entrance aperture
which projects to $0\farcs87 \times 14\farcs0$ on the sky, giving a
resolving power of about 60,000 at 5500 {\AA} and enabling sky
subtraction (typical seeing is $0\farcs6 - 1\farcs2$).  A few 
observations were made with the B5 decker $0\farcs87 \times 3\farcs0$
that does not permit sky subtraction.  It is possible that a few of
those observations suffered from minor moonlight contamination.  The
average exposure was 30 minutes, with some as short as 15 min and
others as long as 45 min, depending on seeing and clouds.

\begin{figure}
\begin{center}
\includegraphics[width=150mm]{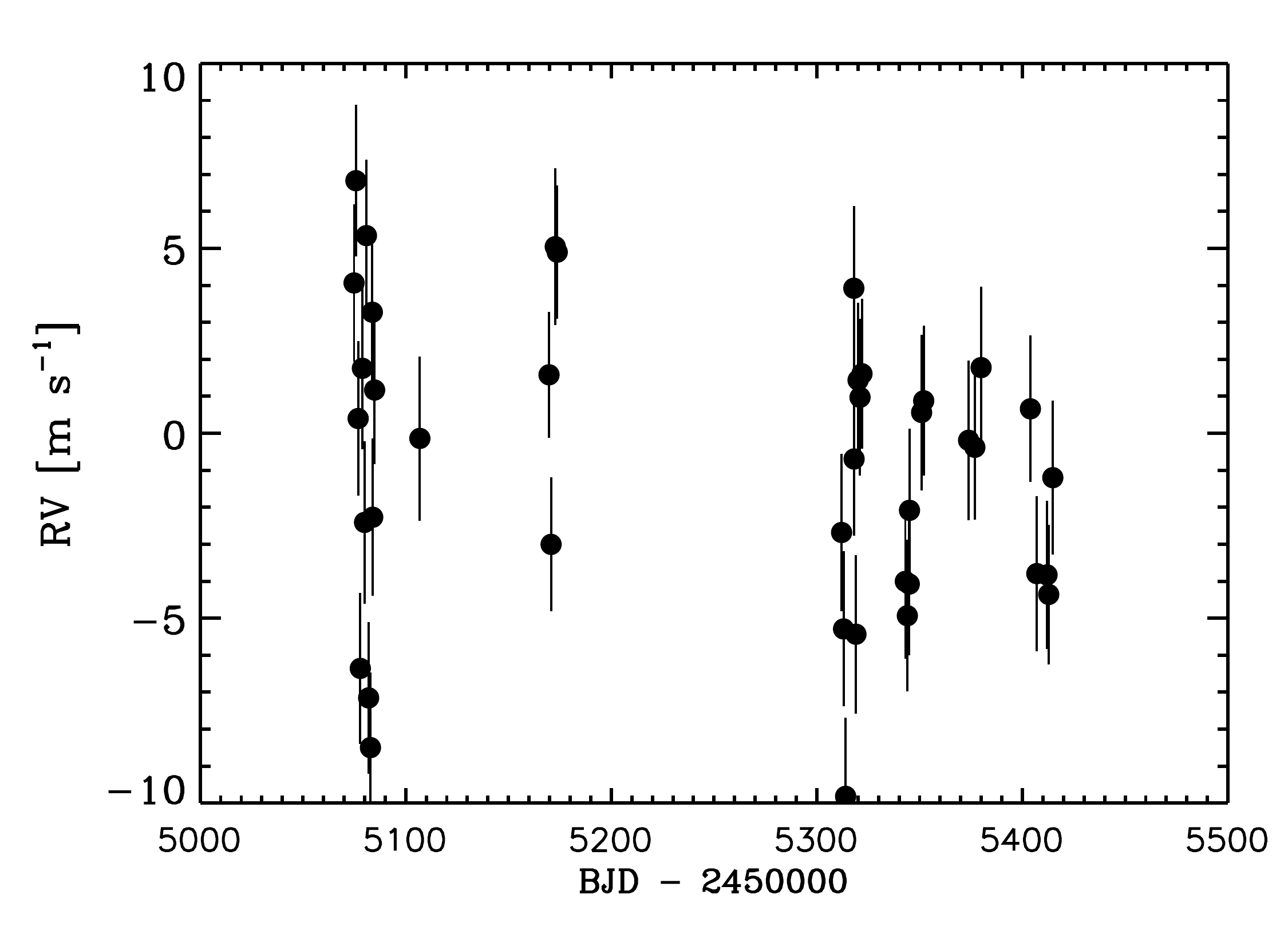}
\end{center}
\caption{Radial velocities derived from HIRES spectra collected in 2009 and 2010 are plotted against time.  Error bars include not only the expected instrumental noise but also a 2 \ms\ jitter to account for variations intrinsic to the star.}
\label{fig:rv_time}
\end{figure}

The raw CCD images were reduced by subtracting an average bias,
subtracting the sky counts at each wavelength just above and below the
stellar spectrum, flat-fielding the spectrum with a 48-exposure sum
from a quartz lamp, and extracting the spectrum with a width that 
includes 99.99\% of the spectrum.  Cosmic rays were removed from the
raw image first.  The pixels typically contained approximately 20000 photons giving
a Poisson-limited signal-to-noise ratio of 140.  We performed the
Doppler analysis with the algorithm of \citet{Johnson09}.  The
internal Doppler errors (the weighted uncertainty in the mean of 400
spectral segments) are typically 1.5-2.0 \ms.  The resulting
velocities are given in Table~\ref{tab:velocities} and shown in Figure~\ref{fig:rv_time} as a function of time.
The error bars include the internal Doppler errors and an assumed jitter of 2 \ms\ (see below), added in quadrature.
The center of mass velocity relative to the solar system barycenter (Gamma Velocity) for \starname\ is
$\gammaVelb$ km s$^{-1}$ (Table~\ref{tab:SystemParams}).  This is an unusually large RV, 
indicative of old disk or even halo membership.  The low metallicity (Section~\ref{sec:spectParam}), $\feh=\fehSME$, magnetic activity (Section~\ref{sec:spectParam}), and asteroseismic age, $\ageKASC$ Gyr (Section~\ref{sec:astero}), also suggest old disk or halo membership. 

Nearby stars with \teff \ near $5600$ K and \logg \ near 4.35 such as \koi \ have been
previously surveyed for precise Doppler work, revealing a noise-like ``jitter'' of $\sim$2.5 \ms \
caused by surface effects including turbulence, spots on the
rotating star, acoustic oscillations, and atmospheric flows associated
with magnetic flux tubes.  While each effect has its own time scale,
it is practical to account for jitter by simply adding it in
quadrature to the internal errors to yield an estimate of the
total uncertainty in the star's velocity.   We have included a jitter
of 2.0 \ms \ in the model fit to all of the data,
photometric and velocities.  

\begin{figure}
\begin{center}
\includegraphics[width=150mm]{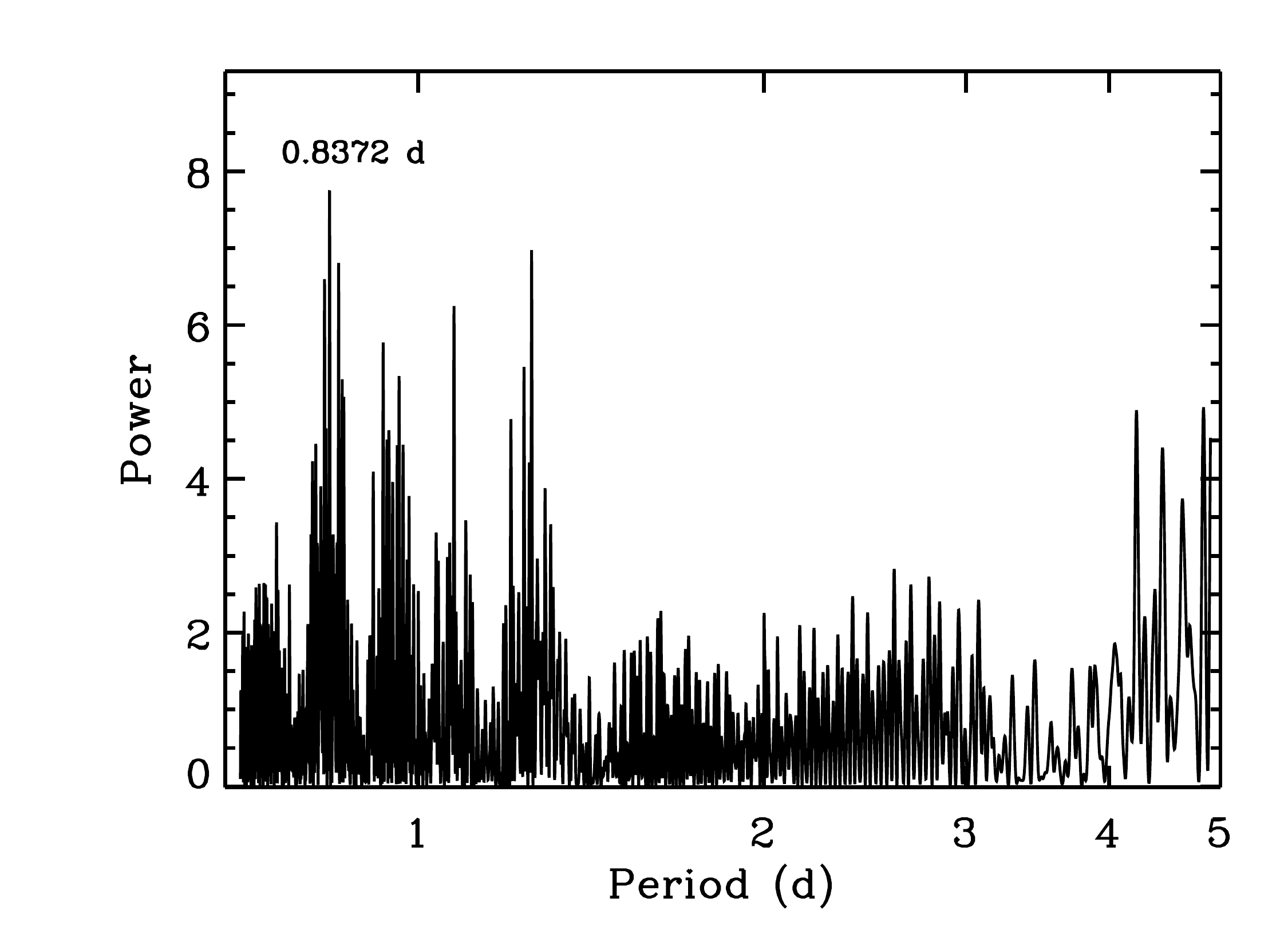}
\end{center}
\caption{A periodogram of the radial velocities shows a peak in the spectral density at the photometric period derived from the transits of \planetb.}
\label{fig:rvPeriodogram}
\end{figure}

A periodogram (Figure~\ref{fig:rvPeriodogram}) of the velocities exhibits a tall peak at a period of
0.837 d, in agreement with the photometric period of the \koib. The coincidence between the transit and RV periods to three significant digits suggests that the RV period is physically related to the transits, as expected if the RV periodicity stems from the reflex motion of the star in response to the gravitational influence of the planet. The periodogram also shows a peak at a period near 1.2 d, which is the alias resulting from the nightly observational cadence.  Similarly, there is another peak near a period of 5 days (off the figure) that is the alias caused by the beating of the 0.827 d period with the 1.0 d cadence of observations.  There is no indication of power at the period of \koic.  

\begin{figure}
\hspace{-50pt}
\includegraphics[height=150mm]{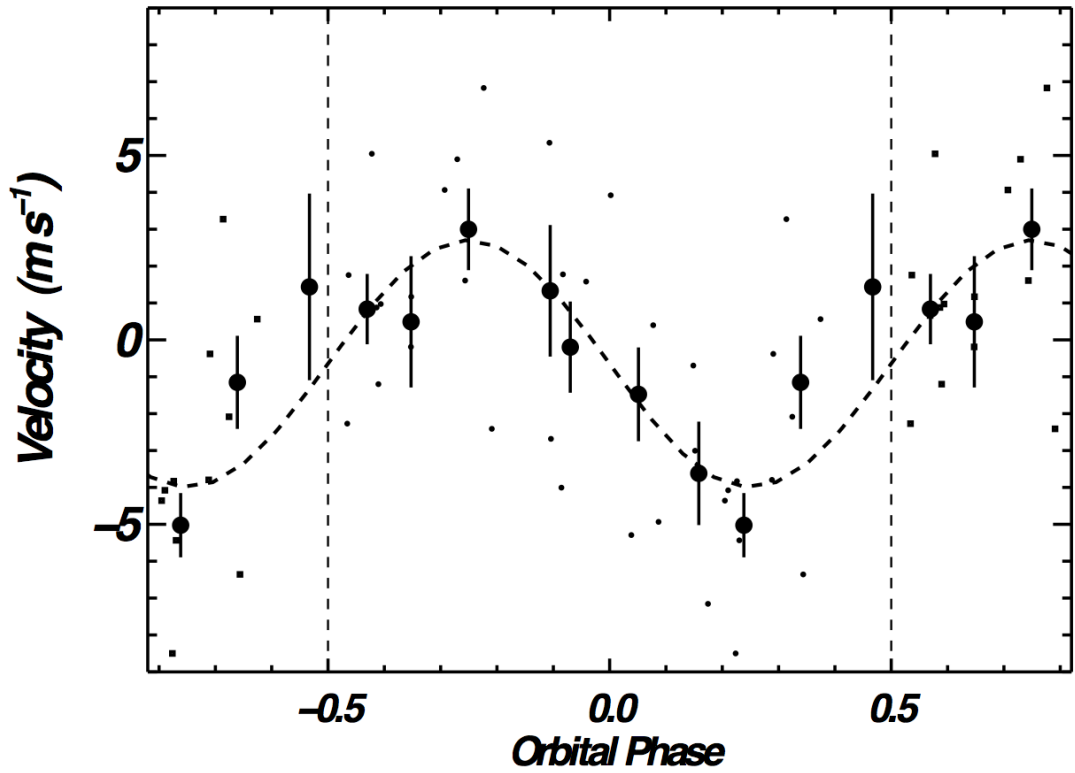}
\caption{Radial velocities versus phase derived from transit photometry of the short-period event.  Both individual velocities are plotted (small circles) as well as averages over 0.1-phase bins (large circles).  The dashed line shows the best-fit circular orbit solution for which there are only two free parameters -- the amplitude $K$ and the zero-point of the velocities.}
\label{fig:rv_phased}
\end{figure}

The velocities phased to the photometric period of \koib\ (Figure~\ref{fig:rv_phased}) show a
clear, continuous, and nearly sinusoidal variation consistent with a nearly circular orbit of
a planetary companion.  The lack of any discontinuities in the phased velocity plot argues against
a background eclipsing binary star as the explanation.  Such a binary
with a period of 0.83 d would have orbital semi-amplitudes of hundreds of
kilometers per sec, so large that the spectral lines would
completely separate from each other, and separate from the lines of the
main star.   Such breaks in the spectral-line blends would cause
discontinuities in the velocity variation, which is not seen here.
Thus, the chance that the 0.83 d periodicity exhibited independently
in both the photometry and velocities might be caused by an eclipsing
binary seems quite remote.

Precision Doppler measurements are used to constrain the mass of \koib\ (\planetb) as discussed in Section~\ref{sec:planet}.  The absence of a Doppler signal for \koic\ is used to compute an upper limit to the mass of this candidate under the planet interpretation.

\subsection{Bisector Analysis}
\label{sec:bisector}

\begin{figure}
\begin{center}
\includegraphics[height=100mm]{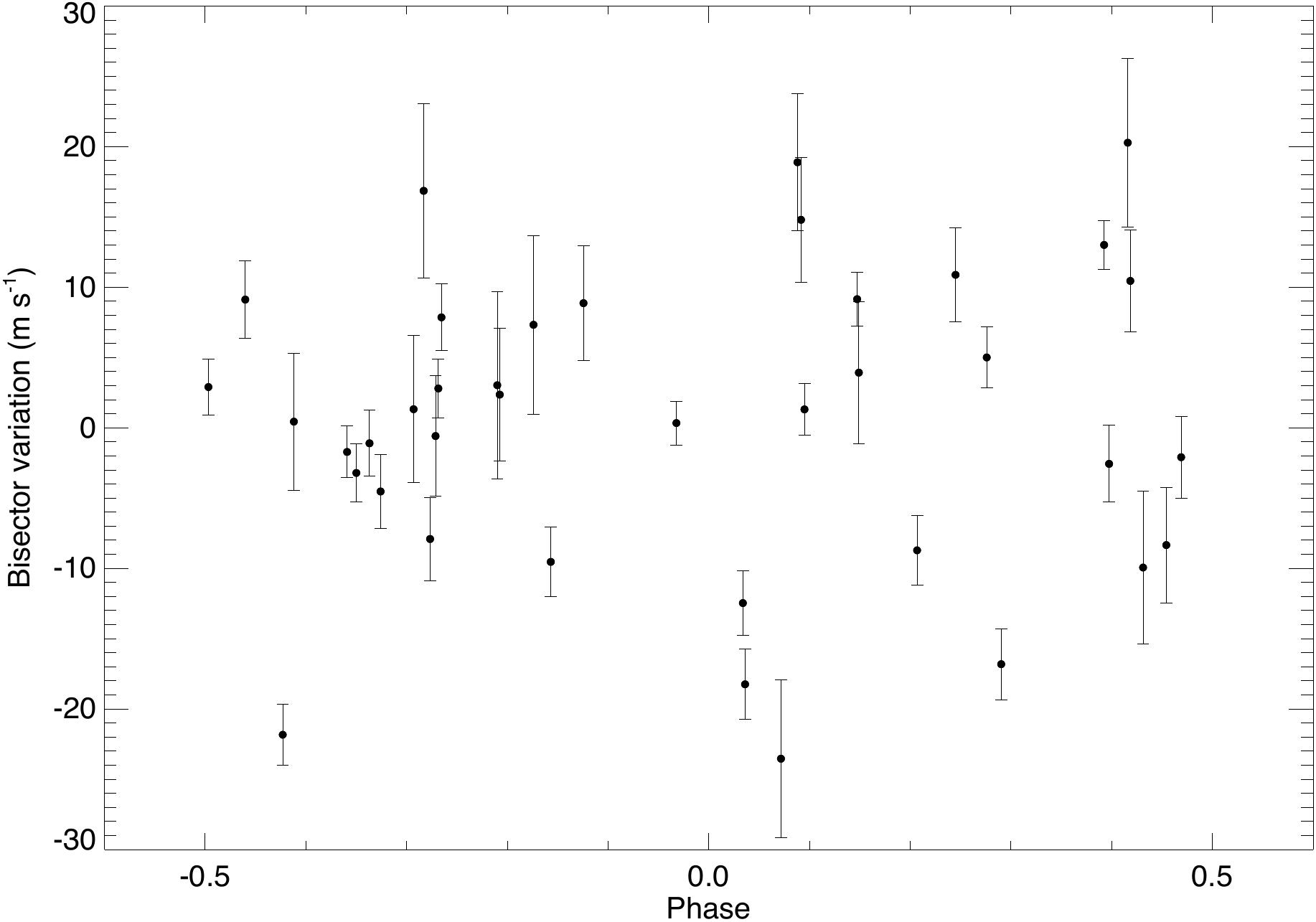}
\end{center}
\caption{Line bisector span measurements folded at the photometric period.}
\label{fig:bisector}
\end{figure}

From the Keck spectra, we computed a mean line profile and the corresponding mean line bisector.  Time-varying line asymmetries are tracked by measuring the bisector spans -- the velocity difference between the top and bottom of the mean line bisector -- for each spectrum \citep{Torres:05}.  When RV variations are the result of a blended spectrum between a star and an eclipsing binary, we expect the bisectors to reveal a phase-modulated line asymmetry \citep{queloz:01, mandushev}. In the case of \planetb, there is no evidence for a correlation between the bisector spans and the RVs which would otherwise argue against the planetary interpretation (see Figure~\ref{fig:bisector}), and similarly for the 45-day signal of \koic.  However, we note that the uncertainties in the bisector span measurements are quite large so that the RMS variation of the bisector spans (10.5 \ms) exceeds the semi-amplitude of the RV variation ($\semiAmpb$ \ms).  Therefore, we do not consider the bisector span measurements to be discriminating in this case.

\section{\blender \ analysis of the \ek\ light curve}
\label{sec:blender}

In this section we examine the possibility that the transit signals
seen in the \ek\ photometry of \starname\ are the result of
contamination of the light of the target by an eclipsing binary along
the same line of sight (`blend'). We consider as potential false
positives physically associated hierarchical triple systems as well as
chance alignments (eclipsing binary in the background or
foreground). We make use of the technique referred to as \blender,
described recently by \cite{Torres:10}, which we apply separately to
each of the signals in \starname\ since each could be due in principle to
a separate blend. Briefly, this technique compares in a $\chi^2$ sense
the observed light curve to a synthetic light curve resulting from
brightness variations of an eclipsing binary being attenuated by the
(typically) brighter star \starname. The parameters of the eclipsing
binary are varied over wide ranges to find all viable blend scenarios
producing a good match to the observations. The properties of each
component of the binary (referred to here as the `secondary' and
`tertiary') are taken from model isochrones \citep{Marigo:08}, and
those of the main star (the `primary') are constrained by the
spectroscopic analysis described earlier. For the technical details of
\blender\ we refer the reader to the previously cited work, as well as
\cite{Torres:04}.

\subsection{\koib \ (\planetb) Signal}
\label{sec:blender-inner}

Given the short period of this signal, we may assume that tidal forces
have circularized the orbits of any potential eclipsing binary
contaminants \citep{Mazeh:08}. We
considered first the case of a hierarchical triple system.
Simulations with \blender\ clearly indicate that such systems in which
the eclipsing binary is composed of two stars provide poor fits to the
\ek\ light curve. We thus rule out this type of blend scenario. If
the eclipsing binary is composed of a planet (i.e., a smaller
tertiary) transiting a star, rather than two stars eclipsing each
other, then it is possible to reproduce the measured light curve, but
only if the secondary has very nearly the same brightness as the
target star itself.  In that case the resulting size of the tertiary
is $\sqrt{2}$ larger than in a model of a single star transited by a
planet.  However, such a bright contaminant would have been evident in
our spectroscopy as a second set of lines, and this case is therefore
also excluded.

\begin{figure}
\epsscale{1.0} 
\plotone{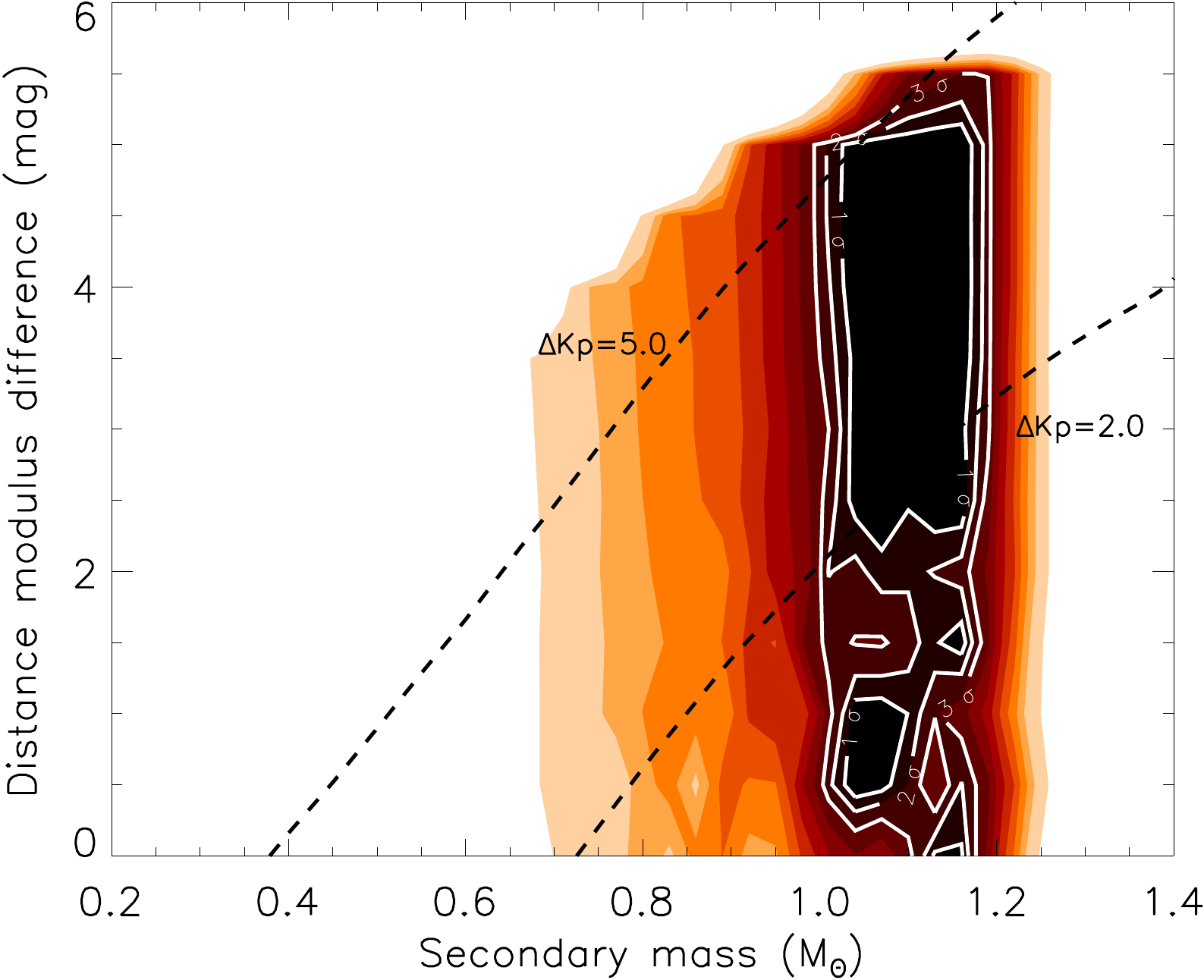}
\figcaption[]{Map of the $\chi^2$ surface corresponding to a grid of
blend models for \koib \ involving background eclipsing systems in
which the tertiary is a (dark) planet, in a circular orbit around the
secondary. Differential extinction is included (see text). The
vertical axis represents a measure of the relative distance between
the background binary and the primary star, which we parametrize here
for convenience in terms of the difference in distance modulus.
Contours are labeled with the $\Delta\chi^2$ difference compared to
the best planet model fit (expressed in units of the significance
level of the difference, $\sigma$). Two dashed lines are also shown
that correspond to equal magnitude difference ($\Delta Kp$) between
the contaminating background binary and the primary star. Kinks in the
contours are simply a result of the discreteness of the
grid.\label{fig:blender1}}
\end{figure}

We next examined the background eclipsing binary scenario, allowing
the relative distance between the binary and the main star to vary
over a wide range. We accounted for absorption from dust along the
line of sight as described by \cite{Torres:10}, adopting a
representative coefficient of differential extinction of $a_v =
0.5$~mag~kpc$^{-1}$.  Interestingly, we found that no combination of
relative distance and stellar properties for the eclipsing binary
(composed in this case of two stars) gives an acceptable fit to the
light curve. The reason is that all such blend configurations lead to
out-of-eclipse brightness changes (ellipsoidal variations) with an
amplitude so large as to be ruled out by the data.
Thus, background blends of
this kind can be confidently ruled out. This result is significant,
because it reduces the overall likelihood of blends for \koib \ 
considerably, as we describe below.  If we allow the tertiary in the
eclipsing pair to be a planet instead of a star (i.e., an object of
smaller radius), then we do find a range of blend scenarios that lead
to acceptable fits to the light curve, which cannot be ruled out a
priori. This is illustrated in Figure~\ref{fig:blender1}, in which we
show contours of equal goodness of fit of the light curve compared to
a standard transit model fit.  \blender \ indicates that these false
positives can be up to about 5 magnitudes fainter than the primary in
the \ek\ band, and that in all cases the secondary star is close
in spectral type (and mass, or color) to the primary, or slightly
earlier. Of these blends, we can further rule out those with
secondaries having $\Delta Kp < 2$, which are bright enough that they
would have been detected spectroscopically. This implicitly places a
lower bound also on the size of the tertiaries
($\sim$3.8\,$R_{\earth}$), as tertiaries smaller than this limit only
give good matches to the light curve if the contaminating star-planet
pair is not too far behind the primary, and is therefore relatively
bright.\footnote{This lower limit of 3.8\,$R_{\earth}$ also excludes
white dwarfs as possible tertiaries.  Additionally, such massive
objects in a tight 0.84-day orbit would lead to very significant
ellipsoidal variation due to tidal distortions induced on the primary
star, which are not seen.}  Those cases would be ruled out
spectroscopically, as mentioned before. The remaining blends shown in
the figure above the line with $\Delta Kp = 2$ must be addressed
statistically.

\begin{figure}
\epsscale{1.0}
\plotone{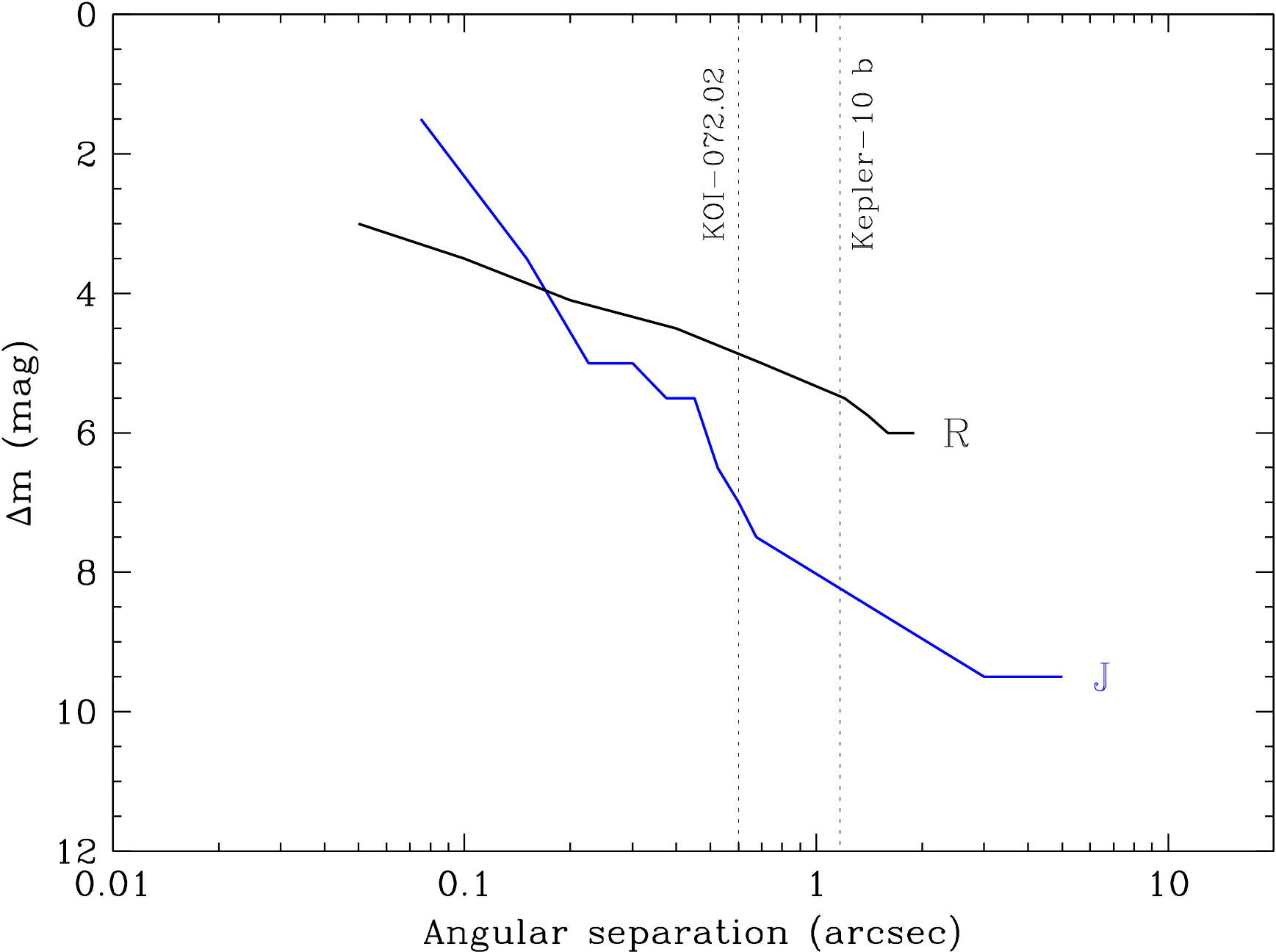}
\figcaption[]{Sensitivity to faint companions near \koi \ from our
imaging observations. Any companions above the curves would be bright
enough to be detected. $J$-band limits are from AO observations at the
Palomar 200-inch telescope, and $R$ is from speckle observations using
the WIYN 3.5\,m telescope. The vertical dotted lines indicate the
3-$\sigma$ confusion radius of 1\farcs17 for \koib \ and
0\farcs60 for \koic, i.e., the maximum angular separation at
which background eclipsing binaries would remain undetected in our
centroid motion analysis.\label{fig:ao}}
\end{figure}

For this we followed closely the methodology applied by
\cite{Torres:10} for the case of Kepler-9d. We computed the mean
density of stars (i.e., background contaminants) in the appropriate
mass range based on Figure~\ref{fig:blender1} in half-magnitude bins,
using the Besan\c{c}on Galactic structure models of \cite{Robin:03},
and we calculated the fraction of these stars that would remain
undetected after our high-resolution imaging observations described
earlier (the constraints from centroid motion analysis are less
stringent; see Figure~\ref{fig:ao}). Some of these stars might be
orbited by (transiting) planets, constituting potential blends. To
estimate how many of these cases one would expect, we adopted the same
frequencies of transiting Jupiters and transiting Neptunes as in
\cite{Torres:10}, based on the results of \cite{Borucki:10}, adjusted
in the case of the Neptune-size planets to account for the lower limit
of $\sim$3.8\,$R_{\earth}$ allowed by \blender\ for the tertiaries.
The outcome of these calculations is presented in
Table~\ref{tab:statistics1}.  The total frequency of false positives
(blend frequency, BF) we expect to find \emph{a priori} for \koib \ 
is ${\rm BF} = 1.4 \times 10^{-8}$. Translating this into a
probability statement for the planet likelihood is difficult, as
argued by \cite{Torres:10}, because it requires knowledge of the rate
of occurrence of super-earth-size planets, a quantity that is not yet
in hand. Nevertheless, following those authors we may express the
false alarm rate for a random candidate star in the \ek\ field
generally as ${\rm FAR} = N_{\rm FP}/(N_{\rm FP} + N_p)$, in which
$N_{\rm FP}$ is the number of false positives and $N_p$ is the unknown
number of planets in the sample. The number of false positives may be
taken to be $N_{\rm FP} = {\rm BF} \times 156,\!097 = 0.0022$, the
product of the blend frequency for \koib \ and the total number of
\ek\ targets \citep{Borucki:10}.

If we were to accept a confidence level of 3$\sigma$ (99.73\%) as
sufficient for validation of a transiting planet candidate
(corresponding to ${\rm FAR} = 2.7 \times 10^{-3}$), then the minimum
number of super-earth-size planets required in order to be able to
claim this level of confidence happens to be $N_p = 1$, according to
the expression above. This value is so small that it gives us high
confidence that the \koib \ signal is not a false positive, but
instead corresponds to a bona-fide super-earth-size planet.
Another way to view this is that the expected number of background
stars capable of producing the observed signal is $\sim8 \times
10^{-6}$ (see Table~\ref{tab:statistics1}, bottom of column 5), so the
probability that what we are seeing is a background blend is $\sim8
\times 10^{-6} \times P_{GP}/P_{SE}$, where $P_{GP}$ and $P_{SE}$ are
the a priori probabilities that giant planets and super-earths occur
with short-period orbits.  Thus, claiming a FAR corresponding to a
3-$\sigma$ detection only requires that the aforementioned probability
ratio does not exceed 300.

It is worth noting that the arguments above allow us to validate
\koib \ \emph{independently} of the detection of the reflex motion
of the star (RV). This is possible in this case because
we are able to rule out, using \blender, all false positives in which
the background eclipsing binary is composed of two stars, which are
predicted to induce ellipsoidal variation at a level that is not
present in the photometry. Were this not the case, the blend frequency
BF would have come out considerably larger, requiring in turn a
significantly higher value for $N_p$.

\subsection{The \koic \ Signal}

A similar \blender\ analysis was performed for \koic. For
circular orbits, hierarchical triple systems are ruled out for the same
reasons as in \koib.  When considering the case of background
eclipsing binaries (stellar tertiaries) with circular orbits, we find
a range of blend scenarios that provide acceptable fits to the light
curve, with secondaries of similar spectral type as the primary.
Unlike the situation for \koib, the longer period of \koic \
(45.3 days) leads to negligible ellipsoidal variation for the
contaminating binaries, and this does not allow these types of blends
to be excluded based on the quality of the fit, as we were able to do
before.  Chance alignment scenarios in which the tertiaries are
planets rather than stars also lead to viable blends that can be up to
4 mag fainter than the target. Those that are brighter than $\Delta Kp
= 2$ can be ruled out because they would have produced a spectroscopic
signature, but fainter ones cannot be ruled out any other way if they
are angularly close enough to the target to be unresolved by our
imaging observations.

However, the longer period of this signal does not justify the
assumption of a circular orbit. Allowing the orbit of a contaminating
binary to have arbitrary eccentricity and also arbitrary orientation
(longitude of periastron) can significantly increase the range of
blends that provide good matches to the \ek\ photometry, both for
the chance alignment case and for physically associated triples. This
is because the orbital speed in an eccentric orbit can be considerably
larger or smaller than in a circular orbit, allowing for blends
involving smaller or larger secondaries than would otherwise be
permitted while still matching the observed duration of the transits,
as described by \cite{Torres:10}.  It also increases the complexity of
the problem, as the space of parameters to be explored is much
larger. Additionally, because the secondaries are now not necessarily
of the same spectral type as the primary star, attention must be paid
to the resulting color of these blends, which could be different from
the measured colors as reported in the \ek\ Input Catalog
\citep{kicLatham}, in which case the blend would be excluded.  Because
these complications require significantly more effort to address, we
are unable to provide sufficient evidence for the planetary nature of
the \koic \ signal at this time based on \blender\ considerations alone, 
and we defer such a study to a forthcoming publication. We note, however, that
the transit duration and period of the two transit signals identified in the light
curve give the same stellar density ($1.142\pm0.092$ g cm$^{-3}$ for \koib\ versus 
$1.147\pm0.096$ g cm$^{-3}$ for \koic) as derived from the transit properties 
using the analytic expression given in Equation 9 of \citet{seager:03}. 
This is a rare coincidence for a blend configuration.

\section{Stellar Characteristics}
\label{sec:star}

\subsection{Spectroscopic Parameters}
\label{sec:spectParam}

We carried out an LTE spectroscopic analysis using the spectral synthesis package SME \citep{Valenti96,
  Valenti05} applied to a high resolution template spectrum from Keck-HIRES of
\starname\ to derive an effective temperature, $\teff  = \teffSMEOrig$
K, surface gravity, $\logg = \loggSMEOrig$ (cgs), metallicity, $\feh = \fehSMEOrig$, 
$\vsini = \vsiniSMEOrig$ \kms, and
the associated error distribution for each of them.  To refine the true parameters of the star, we took a novel path to constrain its surface gravity.  The above effective temperature was used to constrain the fundamental stellar parameters derived via asteroseismic analysis (see Section~\ref{sec:astero}).  The asteroseismology analysis gave $\loggKASC$ which is 0.2 dex lower than the SME value.   The asteroseismology value is likely superior because of the high sensitivity of the acoustic periods to stellar radius.  Still, the asteroseismology result depended on adopting the value of $\teff$ from SME.  We recomputed the SME analysis by freezing (adopting) the seismology value for $\logg$.  This iteration yielded values of $\teff = \teffSME$ K, $\feh = \fehSME$, and rotational $\vsini = \vsiniSME \kms$.   The revised effective temperature was then put back into the asteroseismology calculation to further constrain the stellar radius and gravity.   This iterative process converged quickly, as the \logg\ from seismology yielded an SME value for \teff \ that was only slightly different from the original unconstrained determination.

We also measured the Ca II H\&K emission \citep{Isaacson10}, yielding a Mt. Wilson S value, $S$=0.180 and $\log R'_{HK}$ = -4.89.   Thus \starname\ is a magnetically inactive star, consistent with its low rotational rate, \vsini = 0.5 \kms.   Thus \starname\ appears to be an old (age greater than 5 Gyr) slowly rotating inactive star, slightly above the main sequence.  This is consistent with the age derived from the asteroseismology analysis (Section~\ref{sec:astero}).

\subsection{Asteroseismology and the Fundamental Stellar Properties}
\label{sec:astero}

\begin{figure}
\begin{center}
\includegraphics[height=150mm]{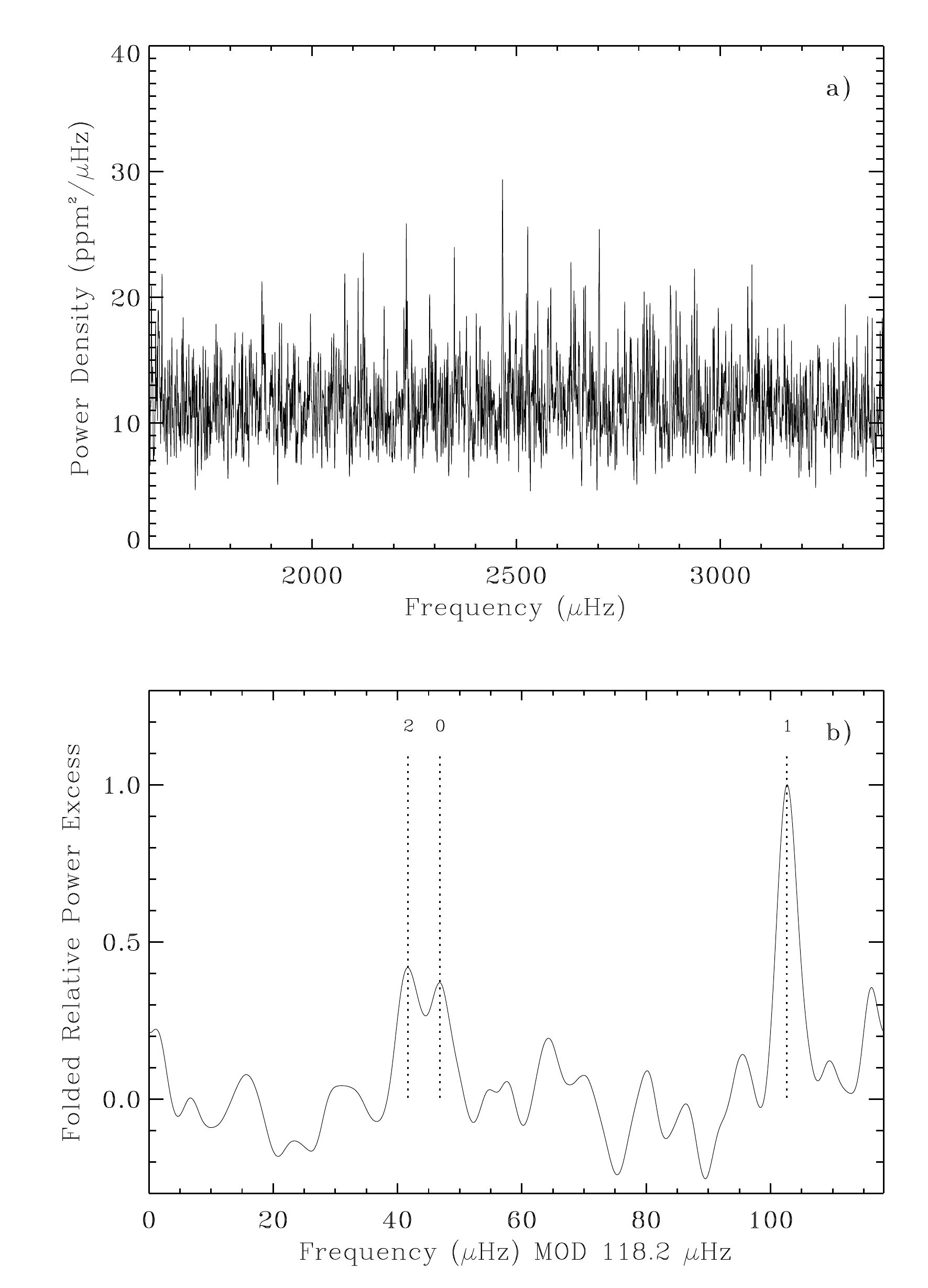}
\end{center}
\caption{Panel a) shows the power-density spectrum, with a 1.6\,$\mu$Hz
smoothing, for short-cadence
observations during the second month of Q2 and all of Q3.
Panel b) shows the spectrum folded at the large separation frequency of 118.2\,$\mu$Hz where the relative power excess is the power minus the noise floor, the result of which is divided by the peak power.
This allows identification of the peaks corresponding to modes of
degree $l = 0$, 1 and 2, as indicated.}
\label{fig:ast1}
\end{figure}

With a magnitude in the \ek\ bandpass of Kp$=\kepmag$, \starname \ presented itself as a promising case for
asteroseismic characterization and was, consequently, placed on the short-cadence target list before it was even identified as a planet candidate.  Figure \ref{fig:ast1}a illustrates the power density spectrum of the short-cadence light curve.
It shows a clear enhancement of power, as expected for solar-like
oscillations, around a frequency of 2500\,$\mu$Hz.
In the spectrum one can identify sequences of approximately uniformly
spaced peaks.
This is in accordance with the asymptotic behavior of high-order
acoustic modes, according to which the cyclic frequencies
$\nu_{nl}$ approximately satisfy
\begin{equation}
\nu_{nl} \simeq \Delta \nu_0 (n + l/2 + \epsilon)
- l (l + 1) D_0
\label{eq:ast1}
\end{equation}
\citep{Vandak1967, Tassou1980},
where $n$ is the radial order and $l$ is the spherical-harmonic degree of the
mode.
The {\it large frequency separation}
$\nu_{nl} - \nu_{n-1\,l} \simeq \Delta \nu_0$ is essentially
given by the inverse sound travel time across a stellar diameter;
it is closely related to the mean stellar density $\rhostar$,
approximately satisfying $\Delta \nu_0 \propto \rhostar^{1/2}$.
For main-sequence stars
$D_0$, giving rise to {\it the small frequency separations}
$\nu_{nl} - \nu_{n-1 \, l+2}$, is largely determined by the
variation of sound speed in the core of the star and hence provides
a measure of the evolutionary state of the star.
Finally, $\epsilon$ is determined by conditions near the surface of the star.\footnote{Owing to the small value of $\vsini$ we do not have to consider rotational effects on the frequencies.}
\citep[For details on the diagnostic potential of solar-like
oscillations, see, for example][]{Christ2004}.
Photometric observations such as those carried out by \ek\
are essentially restricted to degrees $l \le 2$.

The analysis of the observed frequency spectrum largely followed
the procedures used by \citet{Christ2010}, with a pipeline
developed for analysis of the \ek\ p-mode data
\citep{Christetal2008, Huber2009}.
The first step was to carry out a correlation analysis to determine
the large frequency separation $\Delta \nu_0$,
leading to $\Delta \nu_0 = 118.2 \pm 0.2$\,$\mu$Hz.
Using the scaling with the mean density and the corresponding values for
the Sun, this yields a first estimate of 
$\rhostar = 1.080 \pm 0.006\,\gcmc$.
It should be noted, however, that this estimate does not take into account
detailed differences between the structure of the star and the Sun.

\begin{figure}
\begin{center}
\includegraphics[height=125mm]{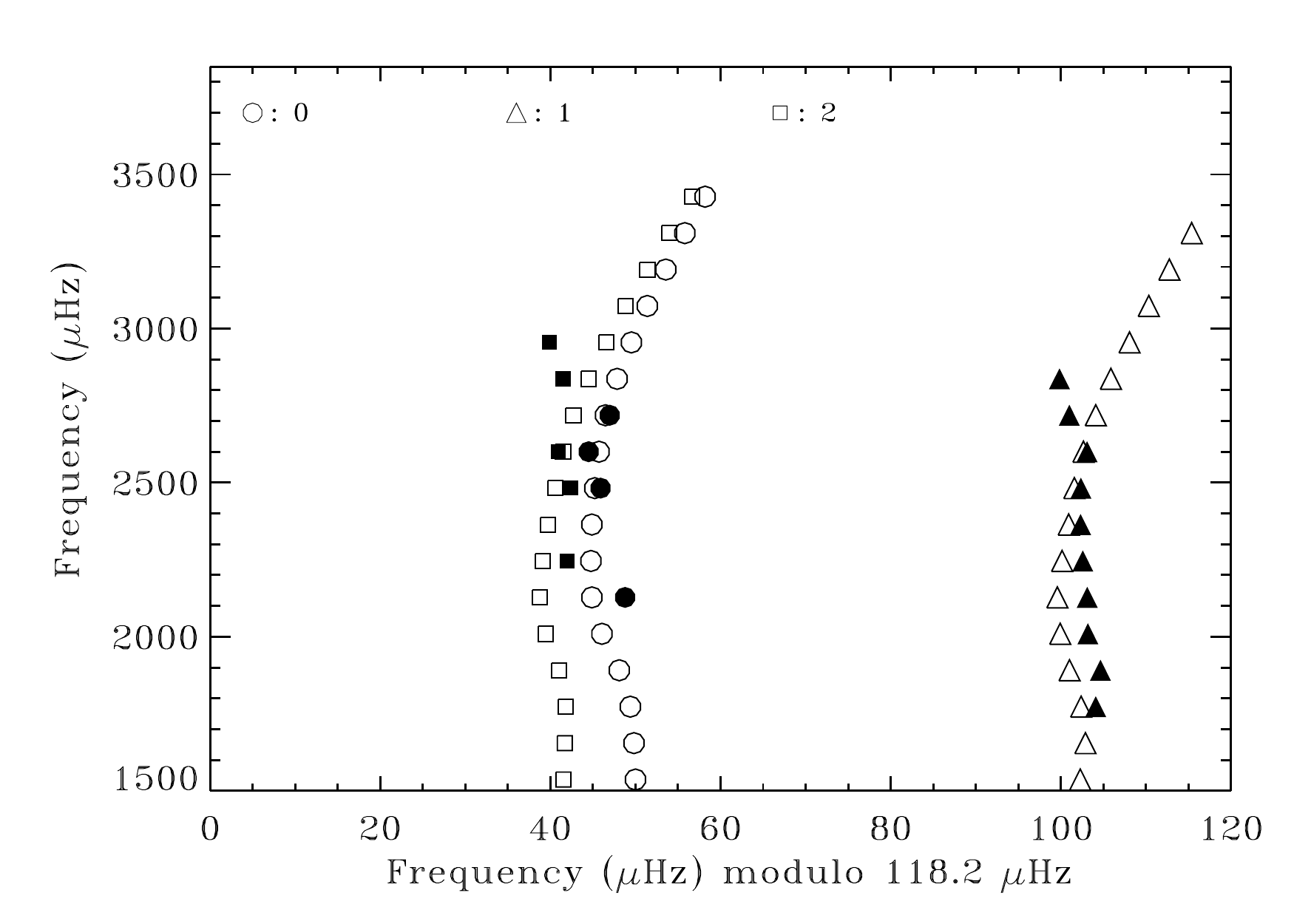}
\end{center}
\caption{\'Echelle diagram \citep[cf.][]{Grec1983} illustrating the
observed frequencies (filled symbols) and the frequencies of one of
the best-fitting
model (open symbols), for modes of degree $l = 0$ (circles), $l = 1$ (triangles)
and $l = 2$ (squares).
The model has a mass of $0.9 \, \msun$, $Z = 0.0144$, $\alpha_{\rm ML} = 1.8$
and an age of 11.6\,Gyr.}
\label{fig:ast2}
\end{figure}

The next step in the analysis was to identify individual modes in the
frequency spectrum.
The detailed structure of the spectrum is illustrated in
Figure~\ref{fig:ast1}b which shows the {\it folded spectrum},
i.e., the sum of the power as a function of the frequency
modulo $\Delta \nu_0$.
As indicated, in accordance with Eq. (\ref{eq:ast1})
there is clearly a pair of closely spaced peaks,
corresponding to $l = 2$ and $0$, as well as a single peak for $l = 1$.
Given this identification, we were able to determine
the individual frequencies of 19 modes.
These are illustrated in an \'echelle diagram \citep[cf.][see below]{Grec1983}
in Fig.~\ref{fig:ast2}.

To determine the stellar properties, we fitted the observed frequencies
to a grid of models.
These were computed using the Aarhus Stellar evolution and pulsation codes
\citep{Christ2008a, Christ2008b}. See also \cite{Christ2010}.
Diffusion and settling of helium and heavy elements were neglected.
The grid consisted of models of mass between 0.8 and $1.1\,\msun$ in
steps of $0.02\,\msun$ and mixing length $\alpha_{\rm ML} = 1.5$, 1.8 and 2.1.
The composition was characterized by heavy-element abundances
$Z = 0.011, 0.0127$ and $0.0144$, corresponding to the observed
${\rm \feh} = -0.15 \pm 0.06$ (see above);
this assumes a solar surface ratio between the heavy-element and hydrogen
abundances $(Z_{\rm s}/X_{\rm s})_\sun = 0.0245$ \citep{Greves1993},
and with the initial hydrogen abundance $X_0$ related to $Z$, from
galactic chemical evolution, through $X_0 = 0.7679 - 3 Z_0$.

For each model in the grid we calculated an evolution track extending
well beyond the end of central hydrogen burning and, for the relevant models
along the track, we computed adiabatic frequencies for modes of
degree $l = 0 - 2$.
The match between the observed and computed frequencies,
$\nu_{nl}^{\rm (obs)}$ and $\nu_{nl}^{\rm (mod)}$,
was characterized by
\begin{equation}
\chi_\nu^2 = {1 \over N - 1} \sum_{nl}
\left( {\nu_{nl}^{\rm (obs)} - \nu_{nl}^{\rm (mod)} \over \sigma_\nu}
\right)^2 \; ,
\label{eq:ast2}
\end{equation}
where $N$ is the number of observed frequencies and $\sigma_\nu$ is the
standard error in the observed frequencies, which we estimated as $1\,\mu$Hz.
In the fit we also considered the observed effective temperature $\teff$
(see above), using as combined measure of the goodness of fit
\begin{equation}
\chi^2 = \chi_\nu^2 + \left({\teff^{\rm (obs)} - \teff^{\rm (mod)}
\over \sigma(\teff)} \right)^2 \; ,
\label{eq:ast3}
\end{equation}
where
$\teff^{\rm (obs)}$ and $\teff^{\rm (mod)}$ are the observed and model values
and $\sigma(\teff)$ is the standard error in $\teff$.

Along each evolution track we determined that model which minimized $\chi^2$,
among the discrete timesteps in the evolution sequence.
The final best model corresponding to the track was obtained by
further minimizing $\chi^2$ to determine $\chi_{\rm min}^2$ for that
track,
interpolating $\teff$ linearly between timesteps and scaling
the frequencies according to $\rhostar^{1/2}$.
Given the resulting model values for all evolution tracks, we determined
the final estimates of the stellar properties as an average
over all tracks, weighted by $\chi_{\rm min}^{-2}$.

A preliminary application of this procedure, using the value
$\teff = \teffSMEOrig$\,K as described in the preceding section,
resulted in an average surface gravity of $\logg = \loggKASC$, substantially different
from the spectroscopically determined value of \loggSMEOrig.
In view of the known problems with the spectroscopic determination
of $\logg$
we repeated the spectroscopic analysis, fixing $\logg$
as $\loggKASC$ in accordance with the asteroseismic inference.
This resulted in $\teff = \teffSME$\,K.
We then repeated the asteroseismic fits, conservatively increasing the uncertainty in \teff\ to 60 K.
This resulted in the final estimates of the stellar parameters presented in Table~\ref{tab:SystemParams}:
$\mstar = \mstarKASC \msun$, $\rstar = \rstarKASC \rsun$, $\rhostar = \rhostarKASC$ g cm$^{-3}$, and age $= \ageKASC$ Gyr.
The distance of $\distance$ pc is computed using the $g$-band SDSS apparent magnitude in the \ek\ Input Catalog and bolometric corrections interpolated form the tables of \cite{girardi}.

To illustrate the quality of the fit, Figure~\ref{fig:ast2}
shows the observed frequencies and the frequencies for the best-fitting model
in an \'echelle diagram, reflecting the asymptotic structure of the
spectrum described by Eq. (\ref{eq:ast1}).
In accordance with this equation, the spectrum has been divided into
segments of length $\Delta \nu_0$ which have been stacked.
Formally this corresponds to reducing the frequencies modulo $\Delta \nu_0$.
The asymptotic behavior is reflected in the nearly vertical columns of
points, corresponding to the different values of the degree.  
Also, it is clear that the model provides a reasonable, although far from
perfect, fit to the observations.
\footnote{It may be noted that, unlike the simple expression (\ref{eq:ast1}),
the small separation between modes with $l = 0$ and 2 depends on frequency.}

\section{Planet Characteristics}
\label{sec:planet}

The physical and orbital properties of both transit signatures are
derived by simultaneously fitting \ek\ photometry and Keck RVs and by
adopting the mean-stellar density of the host star as determined by
asteroseismology. Our system model uses the analytic formalization of \cite{man02}
to fit photometric observations of the transit.
We use the fourth-order non-linear parameterization of limb-darkening also described by \cite{man02} with
coefficients ($c_1=1.086, c_2=-1.366, c_3=1.823, c_4=-0.672$) calculated by \cite{limbdarkening} 
for the \ek\
bandpass.  We account for variability phased to the orbital period by
including the effects of reflected and emitted light from the planet, ellipsoidal variations
due to tidal distortions of the host star and Doppler boosting due to
motion of the star around the center of mass.  For reflected/emitted light,
we assume that the phased lightcurve is reproduced by a Lambertian
reflector scaled by the geometric albedo.  Ellipsoidal variations are modeled
by using the prescription of \cite{pfahl}.  Doppler boosting uses the methodology 
outlined in \cite{dopplerboosting}.  We model the occultation by computing the 
fraction of the planet occulted by the star as a function of the star-planet 
projected distance.  We assume that the planet is a uniformly illuminated disk during occultation.

Our model parameters are the stellar mass and radius (\mstar, \rstar), the planet
mass and radius ($\mpl$, $\rpl$), the orbital inclination
($i$), eccentricity ($e \cos w$, $e \sin w$), the geometric albedo ($A_{g}$),
and the RV amplitude and zero point ($K$, $V_{0}$).  Model fits to the \planetb \
light curve yield an eccentricity that is consistent with zero ($e \cos w = 0.02 \pm 0.10$; $e \sin w = -0.13 \pm 0.20$) which is consistent with our expectations for tidal circularization \citep{Mazeh:08}.  Given the large orbital separation of the outer planet candidate, we can not assume its orbit to be circular based on tidal circularization.  The small predicted RV amplitude prevents a measurement of the eccentricity from existing RV observations.  The duration of the transit for the outer planet candidate is consistent with a circular orbit, but the resulting upper limit is still significant ($e\simeq~0.2$).  For the remainder of our discussion, the models are constrained to zero eccentricity for both \planetb\ and \koic.  The radial velocity variations are modeled by assuming non-interacting (Keplerian) orbits.  If \koic\ were to be confirmed, then the relative inclination between the two orbits is likely less than $20^\circ$, as larger relative inclinations would require a fortuitous alignment of the orbital nodes for both planets to transit \citep{ragozzine}.

We initially fit our observations by fixing \mstar\ and \rstar\ to their asteroseismic
values (see Section~\ref{sec:astero}).  Model parameters are found by chi-squared minimization 
using a Levenberg-Marquardt prescription.  We then use the best-fit values to seed a
Markov Chain Monte Carlo (MCMC) parameter search \citep{ford:05} to fit all model parameters.  We adopt the
uncertainty on the asteroseismic determined mean-stellar density as a prior 
of the stellar mass and radius.   A Gibbs sampler is used with widths
initially defined by uncertainties derived from the diagonals of the constructed
co-variance matrix \citep{ford:06}.  Our Markov-chain contains $194,066$ elements.  We list the median of the
distribution for each model parameter in Table \ref{tab:SystemParams} and the
corresponding $\pm 68.3\%$ credible intervals (akin to a 1$\sigma$ confidence interval) centered on the median.  The resulting properties of \planetb\ are as follows: 
$\mpl=\mplanetb$ \mearth, \rpl=\rplanetb$ \rearth, \rhopl=\rhoplanetb$ g cm$^{-3}$, and a surface gravity ($\loggplanetb$ dex) that is just 2.3 times that of Earth.  Our knowledge of the planet is only as good as our knowledge of the parent star.  Here, the planet radius is determined with a precision of just over 2\% -- comparable to the precision of the stellar radius derived from asteroseismology.  The precision of the planet mass, however, is driven by the low semi-amplitude of the RVs ($\semiAmpb$ \ms) relative to the internal errors (1.5-2 \ms). 

The absence of a statistically significant orbital signature in the RV data at the 45-day period of \koic\ translates to an upper limit for the mass under the planet interpretation.  The best fit to the RVs, constrained by the photometric period and phase, yields a slightly negative semi-amplitude and, consequently, a mathematically valid but physically unrealistic negative planet mass: $\mpl=\mplanetc$ \mearth. Of more relevance is the distribution of the masses returned by the MCMC calculations.  The upper mass limit is taken to be three times the 68.3\% credible interval ($6.5 \mearth$), or 20 $\mearth$.  This is the upper limit reported in Table~\ref{tab:SystemParams}.

\begin{figure}
\begin{center}
\includegraphics[height=140mm]{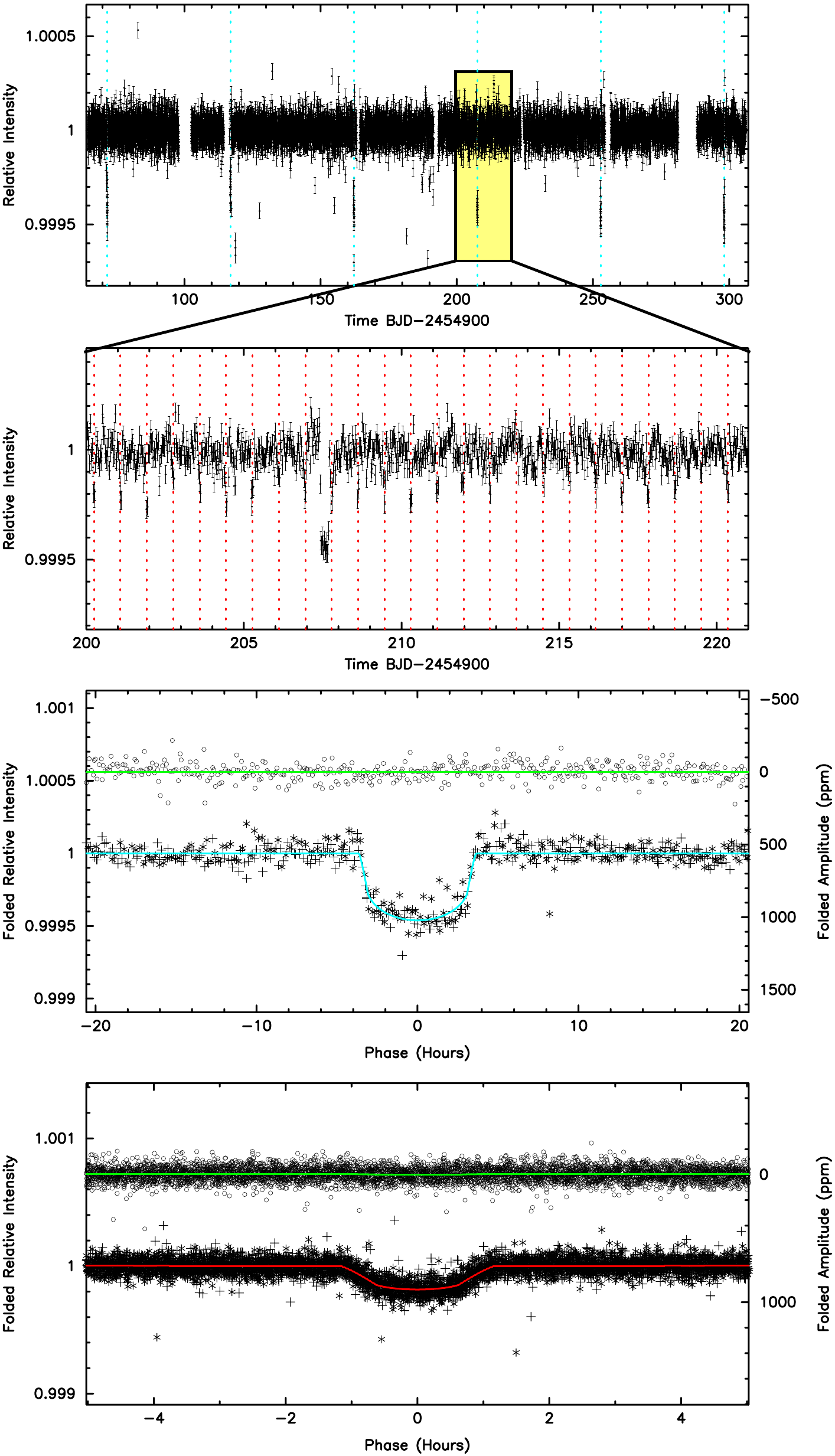}
\end{center}
\caption{The \ek\ photometry and physical models are plotted as a function of both time (upper two panels) and phase (lower two panels).  The transits of \koic\ are highlighted by blue vertical lines in the topmost panel while a cutout (defined by the yellow box) is expanded to show the transits of \planetb\ highlighted by red vertical lines (second from top). The bottom two panels shows the phase-folded light curves centered on phase zero as defined by the central transit time. The modeled light curves are shown as colored lines (blue corresponding to \koic\ and red corresponding to \planetb.  Also shown is a phase cutout of the lightcurve and model (green) centered on phase$=0.5$ where occulations would occur for a circular orbit.  The relative intensity scale for phase$=0$ can be read off the y-axis on the left-hand-side of the plot while the relative intensity scale for phase$=0.5$ can be read off the y-axis on the right-hand-side of the plot.}
\label{fig:modeling}
\end{figure}

\begin{figure}
\begin{center}
\includegraphics[width=150mm]{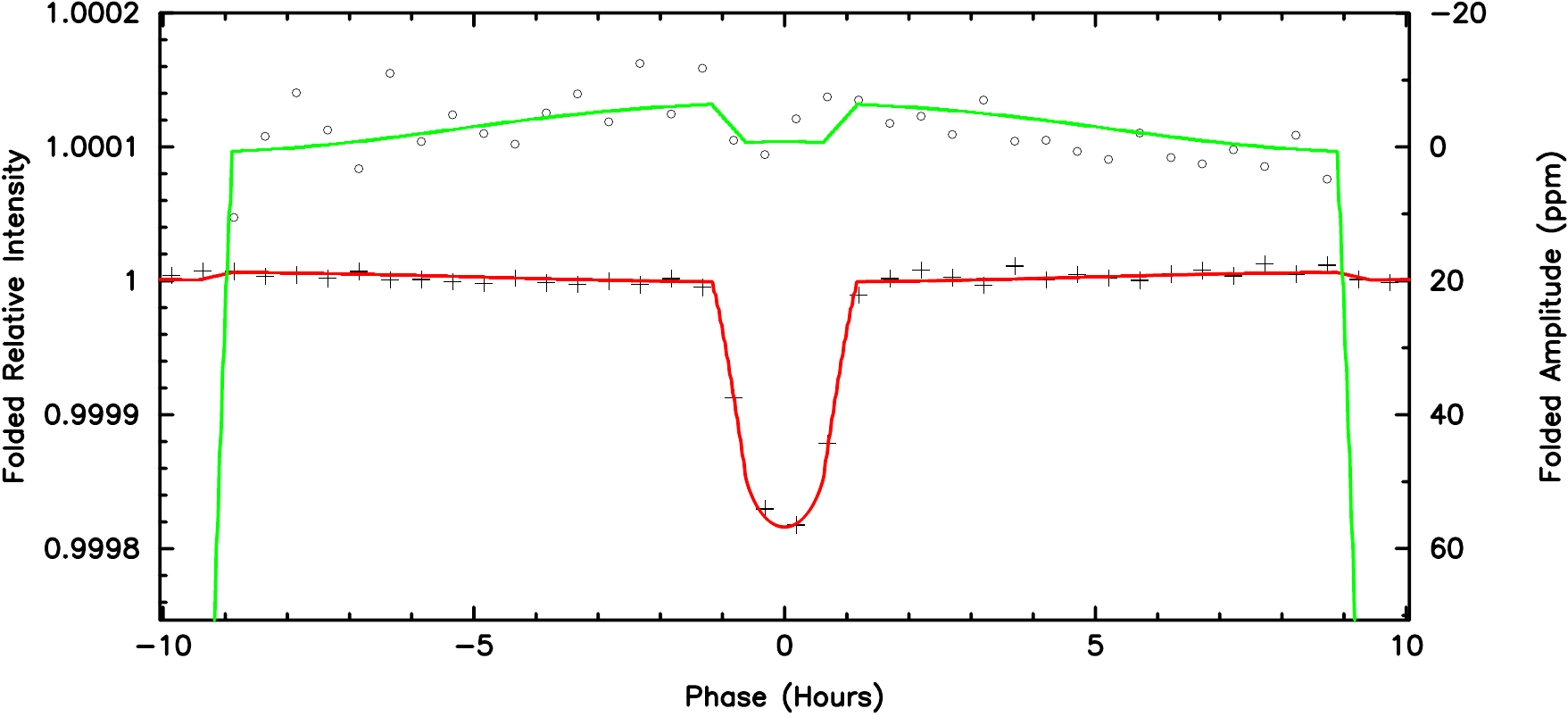}
\end{center}
\caption{The relative intensity scale of the phase-folded light curve of \planetb\ shown in the bottom panel of Figure~\ref{fig:modeling} is expanded to show the phase modulation and marginal occultation required by the model fits. Colors have the same meaning as in Figure~\ref{fig:modeling}.}
\label{fig:phaseCurve}
\end{figure}

The phase-folded light curves together with the model fit are shown in the lower two panels
of Figure~\ref{fig:modeling}.  The best fit for \planetb\ requires a phase modulation with an amplitude 
of $\phaseAmpb$ parts-per-million (ppm) and a $\occultationb$ ppm occultation, both of
which are shown in the scaled and phase-shifted light curve in Figure~\ref{fig:phaseCurve}.  
The modeled RV variations for \planetb\ are shown in Figure~\ref{fig:rv_phased} as a function of 
orbital phase together with the 40 Doppler measurements (small circles) and the same averaged over
0.1-phase bins (large circles).  

The parameters derived from the MCMC analysis are listed in Table~\ref{tab:SystemParams} and discussed in Section~\ref{sec:discussion}. Parameters such as mass, radius, and density for \koic\ are included in the Table for completeness.  However, the reader is cautioned against over-interpretation since the \koic\ transits have not yet been confirmed to arise from a planetary companion.

\section{Statements about transit timing variations}
\label{sec:ttv}

\begin{figure}
\begin{center}
\includegraphics[width=125mm]{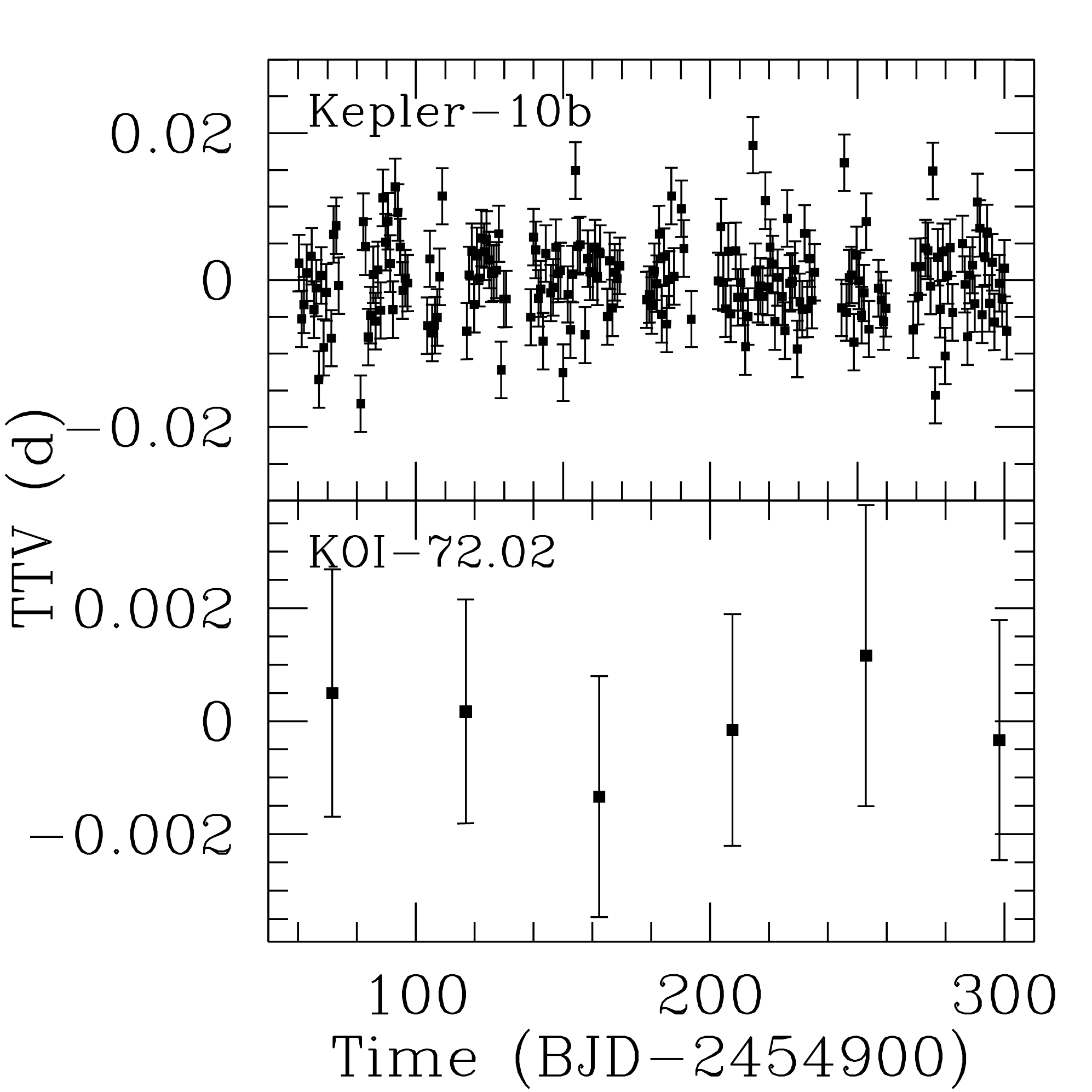}
\end{center}
\caption{The difference between each best-fit transit time
and a linear ephemeris for \planetb\ (top) and \koic\ (bottom).
As expected, we do not detect any statistically significant transit
timing variations.}
\label{fig:ttvTimes}
\end{figure}

To measure the transit times, we generate a template transit shape
based on folding the light curve with the given linear ephemeris.  For
each transit, we estimate the transit time by performing a local
minimization, varying the transit mid-time, the light curve
normalization and the light curve slope (outside of transit), but holding the remaining
parameters (planet-star radius ratio, transit duration, impact parameter and limb darkening)
fixed.  We iterate to improve the
template transit shape used for measuring the transit times.  The
results are shown in Figure~\ref{fig:ttvTimes}.  The true uncertainties may be larger
than the formal uncertainties (indicated by error bars), particularly for \planetb.  
Regardless, we do not detect statistically significant transit timing variations 
for \planetb\ at levels above 0.01 days and for \koic\  at levels above 0.003 days.
Given the masses and periods we measured for both planet candidates, we predicted 
that TTVs would not be detectable, as indeed is the case.

The precision with which the transit times of \planetb \ can be measured is lessened by the small size of the planet.  Coupled with this loss in timing resolution, the period of the planet is also quite short at less than a day.  Thus, the ratio of the period to the timing precision (the signal-to-noise ratio) is only $P/\sigma \simeq (75000 s)/(500 s) = 150$.  For comparison, a Jupiter-size planet ($\sim 1\%$ transit depth) in a 1-week orbit would produce a ratio about 100 times larger. The transit timing variations (TTV) from a nonresonant perturbing planet are essentially independent of the mass of the transiting planet. To summarize, additional non-transiting planets orbiting \starname\ are not well constrained by the measured transit times of \planetb.  Here our RV measurements, which have very good precision of approximately 2.6 \ms, are better suited to detect additional, nontransiting planets in this system.

\begin{figure}
\begin{center}
\includegraphics[width=125mm]{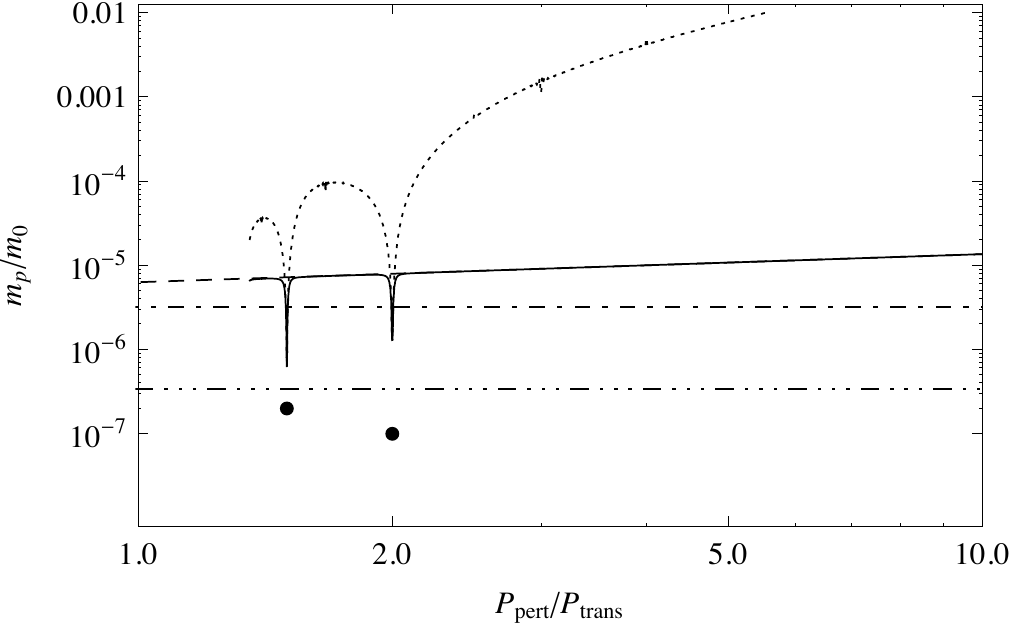}
\end{center}
\caption{Constraints (95\% confidence level) on low-eccentricity secondary planets that are exterior to \planetb\ as a function of the period ratio using the measured transit times and RV data.  The dotted curve shows the limit from a TTV analysis alone from equation (A7) in \cite{agol}.  The dashed line is the expected sensitivity from 40 RV measurements with 2.6 m/s precision calculated using equation (2) from \cite{steffen}.  The solid curve is the overall sensitivity from both RV and TTV measurements (summed in quadrature).  The diamonds are calculations for MMR from equation (33) in \cite{agol}.  The TTV sensitivity curve has been scaled down by $\sqrt{269}$ to represent the improvement in sensitivity due to the number of observed transits.  Finally, the horizontal dot-dashed and triple-dot-dashed lines correspond to the mass of the Earth and the mass of Mars respectively.}
\label{fig:ttvfig}
\end{figure}

The small mass of this planet does provide good sensitivity to resonant perturbing planets where the TTV signal scales with the ratio of planet masses (see equation (33) of \citet{agol}).  Figure \ref{fig:ttvfig} shows the maximum allowed mass of a perturbing planet in an orbit near \planetb \ including both the transit time and RV data.  Near mean-motion resonance, small planets with masses below that of Mars can be excluded.  Away from resonance, the RV data constrain the presence of additional planets to be less than a few Earth masses.

If \koic\ is, indeed, a longer-period planet orbiting \starname, a TTV signal due to the interaction with \planetb\ must exist at some level and it behooves us to consider the expected magnitude of such a signal and whether or not it would be detectable in the data at hand.  The hierarchical architecture limits the mechanisms that are capable of inducing a detectable signal.  Neither changes in the light travel time caused by the displacement of the star due to a hypothetical outer planet, nor an evolving tidal field caused by an eccentric outer planet (see section 4 of \citet{agol}), would be detectable in this system given the measured timing uncertainties.  This statement is true even when one accounts for the factor of $\sqrt{N_{trans}}$ (where $N_{trans}$ is the number of observed transits) statistical improvement in the sensitivity.  A scenario with an outer planet having a large eccentricity would only produce a detectable signal if its mass is comparable to Jupiter and its eccentricity is greater than 0.95 (see equation (25) of \citet{agol}).  Consequently, the absence of a TTV signal in the data does not rule out the planetary interpretation of \koic, given the upper mass limit of 20 \mearth\ (Section~\ref{sec:planet}).

\section{Discussion}
\label{sec:discussion}

\subsection{Composition of \planetb}
\label{sec:composition}

\begin{figure}
\vspace{-20pt}
\begin{center}
\includegraphics[height=130mm]{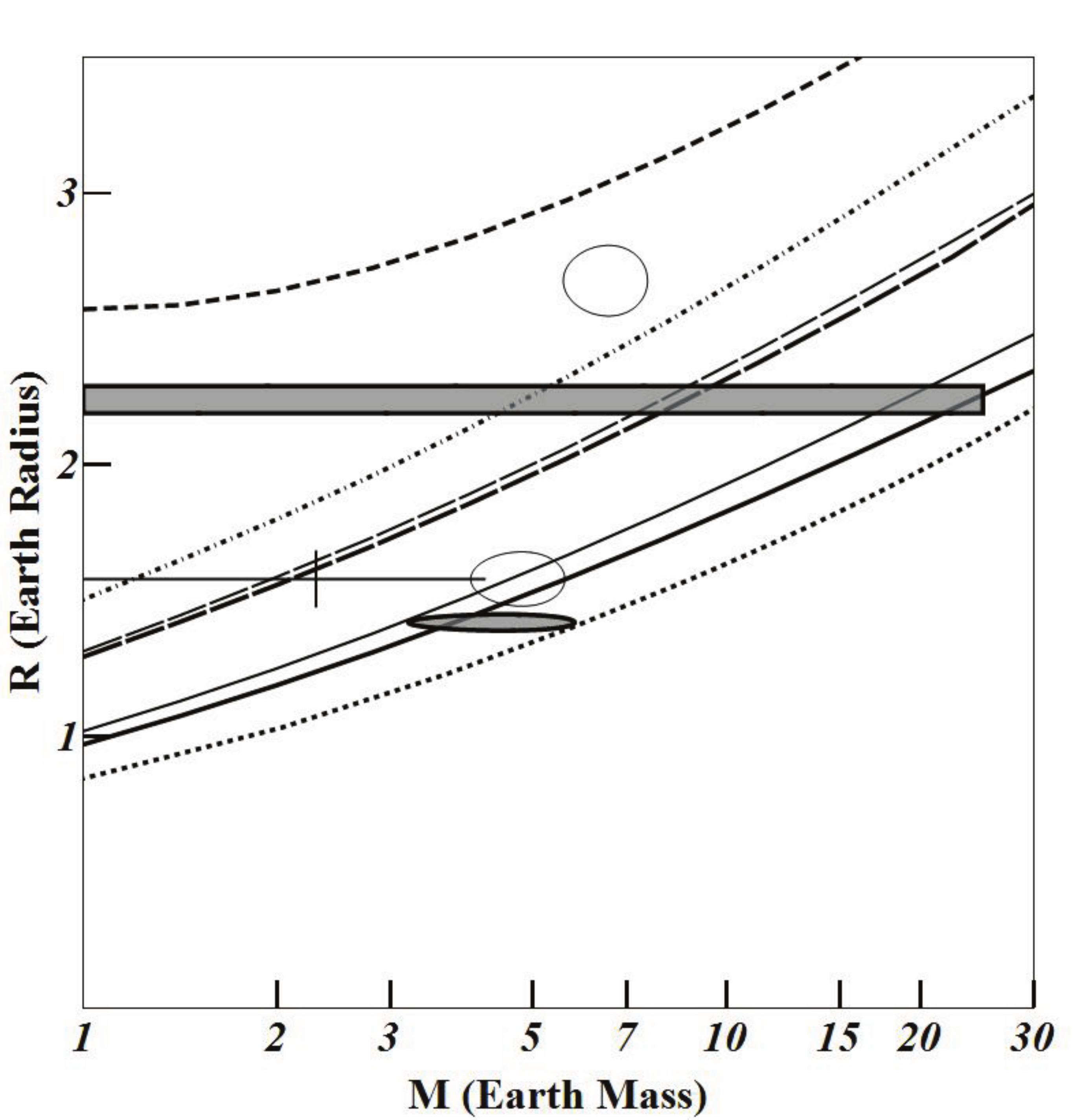}
\end{center}
\caption{\planetb \ and \koic \ on the Mass-Radius diagram. 
\planetb \ is shown with a 1-sigma ellipse at 
$R_p=1.416~R_E$; \koic \ is shown with a 1-sigma band, constrained in 
mass below $\approx$25$M_E$. Two similar planets are shown for
comparison: GJ1214b (1-sigma ellipse top) and CoRoT-7b (1-sigma ellipse below, and solid 
crossed lines for the alternative mass estimate as described in text). 
Theoretical models are shown as curves (from top to bottom): 10\% by 
mass H/He envelope with typical ice giant interior similar to Uranus and 
Neptune (short-dashed line); theoretical pure water object (dot-dashed 
line); 50\% water planets with 34\% silicate mantle and 16\% Fe core 
(thick long-dashed line), or with a low Fe/Si ratio of 44\% mantle and 
6\% Fe core (thin long-dashed line); and earth-like composition with the 
same Fe/Si ratios (thick and thin solid lines). Models are from 
the grid of \cite{zeng}. The dotted curve at the bottom 
corresponds to a maximum Fe core fraction expected from simulations of 
mantle stripping by giant impacts \citep{marcus}.}
\label{fig:massRadius}
\end{figure}

\begin{figure}
\vspace{-130pt}
\begin{center}
\includegraphics[width=150mm]{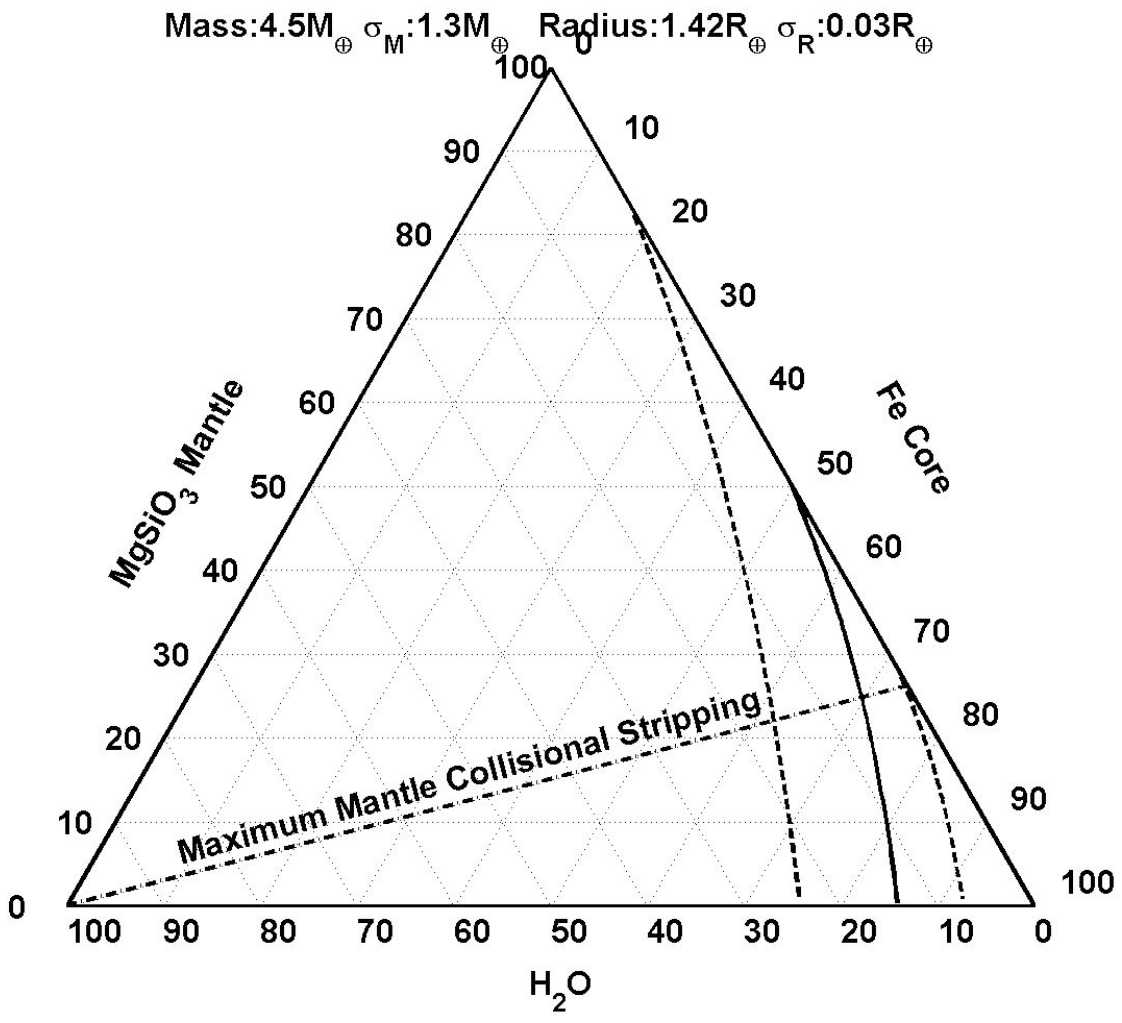}
\end{center}
\caption{The interior structure models for \planetb \ shown on a ternary 
diagram, illustrating the planet's solid nature. Three possible bulk 
materials can mix to determine a planet's radius at a fixed planet mass. 
The model planet radius appears as a curved line; this iso-radius curve 
for \planetb \ (solid line) lies very close to the ``dry side'' of the 
ternary diagram, indicating that most models at this high density 
exclude water as a significant bulk component of the planet's 
composition. The dash-dot line corresponds to a maximum Fe core fraction 
expected from simulations of mantle stripping by giant impacts (\cite{marcus}) .}
\label{fig:ternary}
\end{figure}

\planetb\ is a high-density rocky planet. This conclusion is based on
the comparison of its radius and mass, measured within 1-sigma, with
theoretical calculations of interior structure, as illustrated in
Figures~\ref{fig:massRadius} and \ref{fig:ternary}. The conclusion accounts for known uncertainties in
the theory.  

Figure~\ref{fig:massRadius} is the mass-radius diagram for small planets in units of Earth
mass and radius (after \citet{zeng}). Earth lies on the left
edge, between two models (solid lines) of earth-like composition with a
range of Fe/Si ratios. Uranus and Neptune are out of range, above the
upper right portion of the diagram. \planetb\ (bottom 1-sigma ellipse) and \koic\
(1-sigma band with upper limit in mass at $\approx 25~M_E$) are shown
with models and with two other planets known in this size range:
CoRoT-7b and GJ1214b.  In Figure~\ref{fig:massRadius} the upper 1-sigma ellipses are for GJ 1214b \citep{charbonneau}
and CoRoT-7b (\citet{queloz} for mass; \citet{bruntt}
for radius). The uncertainty in the mass of CoRoT-7b is dependent on the methodology used to correct for the stellar activity signal that dominates the Doppler data.  For example, \citet{queloz} compute a mass of $4.8\pm0.8$ \mearth\ (the value used to construct the error ellipse for CoRoT-7b in Figure~\ref{fig:massRadius}) while \cite{ferraz} obtain $8.5\pm1.5$ \mearth. \cite{hatzes} compute a mass of $6.9\pm1.5$ \mearth, while \cite{pont} obtain $2.3\pm1.8$ \mearth. The mass of CoRoT-7b, provided by \cite{pont}, is shown by straight solid lines in Figure~\ref{fig:massRadius} in order to illustrate the range of mass values present in the literature.
 
Theoretical
calculations in Figure~\ref{fig:massRadius} are shown as curves, which are (from top to
bottom): 10\% by mass H/He envelope with typical ice giant interior
similar to Uranus and Neptune (short-dashed line); theoretical pure
water object (dot-dashed line); 50\% water planets with 34\% silicate
mantle and 16\% Fe core (thick long-dashed line), or with a low Fe/Si
ratio of 44\% mantle and 6\% Fe core (thin long-dashed line; and
earth-like composition with the same Fe/Si ratios (thick and thin solid
lines). These models are from the grid by \cite{zeng}, based
on \cite{valencia:07} with many updates, e.g. new water EOS by
\cite{french}. The Fe/Si ratios chosen in this grid correspond
to the range observed in the Solar System given a stellar Fe/Si range in
the solar neighborhood \citep{grasset}, where the low Fe/Si ratio
corresponds to the lowest values measured in small bodies in the outer
Solar System, namely Ganymede. Note that the hypothetical 100\% water
and H/He envelope model curves are shown for illustration only; given
the extremely high equilibrium temperature of \planetb\ an extended hot
atmosphere of such volatiles has to be accounted for separately, but
\planetb\ is too dense for these scenarios.  

The interior structure of GJ 1214b has been modeled as an H/He/H$_2$O planet with a rocky core \citep{nettelmann}. Indeed, it lies between the models of ice giants similar to Neptune and Uranus and the models of a 50\% water planet.  And while the \citet{queloz} mass and radius point to a rocky composition, the lower mass of \citet{pont} marginally favors a water/ice composition. The properties of \planetb\ together with their uncertainties clearly indicate a high-density, rocky planet. 

The dotted curve at the bottom of Figure~\ref{fig:massRadius} is the envelope
corresponding to a maximum Fe core fraction expected from simulations of
mantle stripping by giant impacts \citep{marcus}. These
simulations and planet formation scenarios indicate that pure Fe core
objects cannot form in the mass range we consider. At the mass of
\planetb\ these simulations predict that the maximum attainable Fe core
is about 75\% by mass. For comparison, Mercury has a $\approx$70\% Fe core.

Theoretical calculations of the interiors of high-density planets suffer
from a number of uncertainties in equations of states, high-pressure
phases of materials inaccessible to lab experiments, cooling and
differentiation histories, etc. There is consensus in the literature
\citep{valencia:06,valencia:07,fortney,seager:07,grasset}
regarding the general results. However, the
theoretical uncertainties are compounded by a degeneracy between only 2
observables (radius and mass), and 3 or 4 distinct types of bulk
materials: Fe core, mantle (silicates, etc.), water, and hydrogen/helium
gas (see \citet{valencia:07,rogers} for details).

At very high densities the radius of solid planets is constrained by the
lack of bulk materials with density and compression properties above
that of Fe and Fe alloys (see lower-envelope dotted curve in Figure~\ref{fig:massRadius}).
Planets with high densities close to that envelope are likely composed
predominantly of silicates and Fe; the degeneracy is lifted as indicated
by the curvature of the iso-radius curves when they approach
asymptotically the ``dry'' right-hand side of the ternary diagram in
Figure~\ref{fig:ternary}.

Ternary diagrams were introduced to studies of solid planet structure 
by \cite{valencia:07}.  The three axes of the ternary diagram represent the core, mantle, and
ice fraction of a planet of a given mass (in this case, the mass derived for \planetb).  The axes are read by following lines parallel to the edges so that the three mass fractions sum to unity.  Although there are 6 radiants emerging from each point in the grid, only three will result in mass fractions that add to unity.  Different combinations of core, mantle, and ice fraction yield a planet radius that can be computed with theoretical models constrained by a given mass.  Radius is not unique to a specific combination of mass fractions.  The solid line in Figure~\ref{fig:ternary} is an iso-radius.  It shows the possible combinations that all yield the derived radius of \planetb.  The dotted lines provide bounds on the domain captured by the one-sigma errors in planet radius. A ternary diagram represents a cross-section at a given mass.  However, mass-dependent uncertainties are included in the error bars represented by the dotted line. The iso-radius together with the one-sigma error bars are a good
representation of the possible bulk compositions for a given planet. 

The degeneracy in bulk composition discussed above is best illustrated in
Figure~\ref{fig:ternary}, as a planet defined by a mass and radius is represented by a
curve (solid) that can span a range of iron:mantle:water fractions. This
degeneracy is practically lifted only for very high density planets near
the ``dry'' right-hand side, close to the 100\% iron core vertex, because:
(1) the iso-radius curves bend as they approach the 0\% water level, and
(2) the H$_2$O and OH molecules are able to be incorporated inside
high-pressure silicate phases without changing their EOS much. For
\planetb, the combination of very high density and constraints from
mantle-stripping simulations (dash-dot line), restrict its bulk
compositions to dry rocky iron-core-dominated interiors similar to
Mercury. On the other hand, if the observational derivation of the mass
of \planetb\ were significantly overestimated (by more than 1-sigma),
the planet could contain significant amount of water in its interior.

In conclusion, within 1-to-2-sigma in its derived radius and mass, the
planet \planetb\ is a dry rocky planet with high Fe content. Its high
density does not violate the prediction for maximum mantle stripping
during planet formation \citep{marcus}.

\subsection{The Phase Curve of \planetb}
\label{sec:phaseCurve}

The phase curve amplitude of \planetb\ is $\phaseAmpb$ ppm (See Table~\ref{tab:SystemParams}).  If
due to scattered light alone, this corresponds to a \ek\ bandpass
effective geometric albedo of 0.68.
The occultation depth is $\occultationb$ ppm which corresponds to an
effective geometric albedo of $\geomAlbedob$. This is an unusually high 
albedo. The only known solar system bodies that
are so bright are Venus (due to photochemically induced hazes) and
Saturn’s icy moon Enceladus (coated with fresh ice). \planetb\ is
likely too hot for any hazes and is certainly too hot for any ice.
Another possibility is that \planetb\ has silicate clouds, but they
would have to be completely covering the planet's day side and have a large
particle size in order to provide the required reflectivity (see, for example,
\citet{seager:00}).

We prefer an interpretation that the phase curve is dominated by a  
thermal radiation change from the planet's day to night side. This  
case would be reminiscent of hot Jupiters, which have a hotter day  
side than night side. In the case of the \ek\ bandpass, the  
temperature difference need not be too extreme, as long as the planet  
is hot enough on the day side (as is the case for \planetb \ with a equilibrium  
temperature of $\teqb$ K). This is because the \ek\ bandpass is in  
the optical, and so the contribution of thermal radiation drops off  
rapidly with decreasing temperature (i.e. day-to-night side). In other words,
a passband on the Wien side of the black body curve will capture significantly different
fluxes with just a small change in temperature.  An example of this is
the phase curve of HAT-P-7 \cite{borucki:09,welsh}. 

An intriguing possibility is that \planetb\ has no atmosphere at all,
having lost it over time due to atmospheric erosion. One test is to
search for a hotspot directly at the substellar point, since no
atmospheric winds would be available to move the hotspot off center
\citep{seager:09}.  Further \ek\ data, in particular a
confirmation and robust measurement of the occultation depth, of
\planetb\ will help with any interpretation.

Asteroseismic measurements of the host star \starname\ indicate that the
planetary system is very old, allowing for more than 11 Gyr of
evaporation and mass loss from the surface of \planetb. Assuming, as
is common, that the planet arrived at its present orbit within about 100
Myr of formation, the relevant evaporation time scales are those of water
steam or a silicate surface. Any H/He envelope would have evaporated too
quickly (at $\approx$ 10$^{11}$ g s$^{-1}$) to affect the further evolution of
the planet, while the evaporation of the silicate mantle is too slow to
remove more than 50\% of the planet's original mass \citep{valencia:09}.
Therefore, it is difficult to establish if \planetb\ is the
remnant core of a water planet or an ice giant planet, though it is
probable that it must have lost a significant fraction of its mass.

\section{Summary}
\label{sec:summary}

NASA's \ek\ Mission collects transit photometry from a spaceborne Schmidt camera to detect and characterize extrasolar planets with the goal of determining the frequency of earth-size planets in or near the habitable zone of Sun-like stars.  Now in its second year of operation, the mission has reached an important milestone toward meeting that goal, namely the discovery, reported herein, of its first rocky planet, \planetb. This planet was identified via transit photometry. The analyses described here are based on $\sim 8$ months of \ek\ 29.4-minute cadence data acquired between 02 May 2009 and 09 January 2010.  The target was also observed at a higher 1-minute cadence from 21 July 2009 to 19 August 2009 and 18 September 2009 and 09 January 2010. 

Two distinct sets of transit events were detected in the lightcurve of \starname \ constructed from $\sim 8$ months of \ek\ photometry: 1) a $\depthb$ ppm dimming lasting $\durationb$ hours with transit ephemeris of $T [BJD] =\epochb + N*\periodb$ days and 2) a longer-period event described by a $\depthc$ ppm dimming lasting $\durationc$ hours and an ephemeris $T [BJD] =\epochc + N*\periodc$ days.  Statistical tests on the photometric and individual pixel flux time series of \starname \ established the viability of the planet candidates.  For example, comparison of the flux-weighted photocenter during transit and outside of transit revealed no deviation consistent with an eclipse event associated with one of the nearby stars that might be diluting the lightcurve of \starname.  Clean statistics triggered a battery of ground-based follow-up observations. 

High resolution reconnaissance spectroscopy was used to verify the effective temperature and surface gravity as well as rule out obvious eclipsing binaries masquerading as planets by way of moderate-precision RVs.  High spatial resolution imaging (AO and Speckle) was acquired to identify faint, nearby stars that should be considered in the photocenter analysis as potential background eclipsing binaries.  In the case of \starname, no additional stars were identified.  Forty precision Doppler measurements were acquired with the Keck 10-meter telescope between May and August 2009.  These measurements confirmed the planetary nature of the short-period transit event.  The photometric period was clearly seen in a periodigram of the velocities, and the variations are phased as expected given the transit epochs.  With a semi-amplitude of just $\semiAmpb$ \ms, the Doppler measurements suggest a planetary mass for this companion.  No significant signal was detected in the measurements at the photometric period of the outer candidate, \koic.

Knowledge of the planet is only as good as our knowledge of the star it orbits.  Matching the reconnaissance spectroscopy of \starname\ to a library of synthetic spectra yielded $\teff=\teffRecon$ K, $\logg=\loggRecon$, $\feh=\fehRecon$, and $\vsini=\vsiniRecon$ \kms.  Full spectral synthesis using HIRES echelle data without the iodine cell yielded $\teff=\teffSMEOrig$ K, $\logg=\loggSMEOrig$, $\feh=\fehSMEOrig$, and $\vsini=\vsiniSMEOrig$ \kms.  Because the parent star is relatively bright (Kp $= \kepmag$) and, hence, amenable to asteroseismic analysis, \ek\ photometry was also collected at 1-minute cadence for $\sim 5$ months from which we detected 19 distinct pulsation frequencies. Modeling of these frequencies resulted in precise knowledge of the fundamental stellar parameters.  The process was iterated once in that the asteroseismic analysis yielded a surface gravity of $\logg=\loggKASC$ which was then fixed in the spectral synthesis analysis to yield an improved effective temperature, $\teff=\teffSME$ K, that was then fed back to the asteroseismic analysis.  The result is that \starname\ is a relatively old (\ageKASC \ Gyr) but otherwise Sun-like Main Sequence star with $\teff=\teffSME$ K, $\mstar=\mstarKASC$ \msun, and $\rstar=\rstarKASC$ \rsun.  

Physical models, constrained by the asteroseismology-derived stellar parameters, were simultaneously fit to the transit light curves and the precision Doppler measurements.  Modeling produced tight constraints on the properties of \planetb: $\mpl=\mplanetb$ \mearth, \rpl=\rplanetb$ \rearth, and \rhopl=\rhoplanetb$ g cm$^{-3}$.  Evaluation of these properties within a theoretical framework allowed us to draw conclusions about the planet's composition.  Within 1-2$\sigma$ of the derived mass and radius, \planetb\ is a dry, rocky planet with high Fe content.  Its high density does not violate predictions for maximum mantle stripping during planet formation. 

\acknowledgements 

The authors would like to thank Carly Chubak for computing the barycentric radial velocity of \starname and Frederic Pont for his careful reading and thoughtful comments, providing us with feedback in record time. JCD acknowledges support from The National Center for Atmospheric Research which is sponsored by the National Science Foundation. Funding for this Discovery mission is provided by NASA's Science Mission Directorate.

\clearpage





\begin{deluxetable}{lccc}
\tabletypesize{\scriptsize}
\tablewidth{0pc}
\tablecaption{Observed flux-weighted centroid shifts for \koib.\label{tab:measured_centroids_01}}
\tablehead{
\colhead{} &
\colhead{Q1} &
\colhead{Q2} &
\colhead{Q3}
}
\startdata
$\Delta R$ & \phs$6.07\times10^{-5} \pm2.42\times10^{-5}$ & \phs$7.36\times10^{-6} \pm2.40\times10^{-5}$ & $-4.91\times10^{-7} \pm2.42\times10^{-5}$  \\
$\Delta C$ & \phs$6.96\times10^{-5} \pm1.59\times10^{-5}$  & $-1.12\times10^{-5} \pm1.25\times10^{-5}$ & $-8.75\times10^{-6} \pm1.59\times10^{-5}$  \\
$D$ & \phs$9.23\times10^{-5} \pm1.99\times10^{-5}$  & \phs$1.34\times10^{-5} \pm9.25\times10^{-5}$ & \phs$8.77\times10^{-6} \pm1.59\times10^{-5}$  \\
$D/\sigma$ & $4.64$  & $0.79$ & $0.55$  \\ [-1.5ex]
\enddata
\end{deluxetable}

\begin{deluxetable}{lccc}
\tabletypesize{\scriptsize}
\tablewidth{0pc}
\tablecaption{Observed flux-weighted centroid shifts for \koic.\label{tab:measured_centroids_02}}
\tablehead{
\colhead{} &
\colhead{Q1} &
\colhead{Q2} &
\colhead{Q3}
}
\startdata
$\Delta R$ & \phs$2.70\times10^{-4} \pm4.08\times10^{-5}$ & $-1.90\times10^{-4} \pm5.35\times10^{-5}$ & \phs$3.14\times10^{-5} \pm3.05\times10^{-5}$  \\
$\Delta C$ & \phs$7.58\times10^{-5} \pm2.87\times10^{-5}$  & \phs$1.91\times10^{-4} \pm4.03\times10^{-5}$ & $-3.84\times10^{-5} \pm2.45\times10^{-5}$  \\
$D$ & \phs$2.81\times10^{-4} \pm4.01\times10^{-5}$  & \phs$2.69\times10^{-4} \pm4.73\times10^{-5}$ & \phs$4.96\times10^{-5} \pm2.70\times10^{-5}$  \\
$D/\sigma$ & $7.01$  & $5.68$ & $1.83$  \\ [-1.5ex]
\enddata
\end{deluxetable}

\begin{deluxetable}{lccc}
\tabletypesize{\scriptsize}
\tablewidth{0pc}
\tablecaption{PRF-fit centroid shifts for \koic \ without saturated pixels.\label{tab:prf_centroids_02}}
\tablehead{
\colhead{} &
\colhead{Q1} &
\colhead{Q2} &
\colhead{Q3}
}\startdata
$\Delta R$ & \phs$3.05\times10^{-4} \pm7\times10^{-5}$ & $-1.82\times10^{-4} \pm7\times10^{-5}$ & \phs$5.50\times10^{-5} \pm7\times10^{-5}$  \\
$\Delta C$ & \phs$1.72\times10^{-4} \pm7\times10^{-5}$  & $-1.84\times10^{-4} \pm7\times10^{-5}$ & \phs$4.39\times10^{-5} \pm7\times10^{-5}$  \\
$D$ & \phs$3.50\times10^{-4} \pm7\times10^{-5}$  & \phs$2.59\times10^{-4} \pm7\times10^{-5}$ & \phs$7.05\times10^{-5} \pm7\times10^{-5}$  \\
$D/\sigma$ & $4.9$  & $3.7$ & $1.0$  \\ [-1.5ex]
\enddata
\end{deluxetable}

\begin{deluxetable}{lccc|lccc}
\tabletypesize{\scriptsize}
\tablewidth{0pc}
\tablecaption{Modeled centroid shifts due to transits on the known stars in the Q3 aperture with depths that
reproduce the observed Q3 depth.\label{tab:modeled_offsets}}
\tablehead{
\colhead{Object} &
\colhead{Modeled depth} &
\colhead{Modeled $D$} &
\colhead{$D/\sigma$} &
\colhead{Object} &
\colhead{Modeled depth} &
\colhead{Modeled $D$} &
\colhead{$D/\sigma$}
}\startdata
\koib & $1.67\times10^{-4} $ & $6.10\times10^{-6} $ & $0.384$ & \koic & $5.52\times10^{-4} $ & $2.02\times10^{-5} $ & $0.747$ \\
11904143 & $1.90\times10^{-1}$ & $5.69\times10^{-4} $ &  $35.8$ & 11904143 & $6.30\times10^{-1}$ & $1.89\times10^{-3} $ &  $69.7$   \\
11904165 & $6.31\times10^{-2}$ & $3.53\times10^{-4} $ &  $22.2$ & 11904165 & $2.09\times10^{-1}$ & $1.17\times10^{-3} $ &  $43.3$  \\
11904167 & $1.17\times10^{-1}$ & $4.43\times10^{-4} $ &  $27.9$ & 11904167 & $3.88\times10^{-1}$ & $1.47\times10^{-3} $ &  $54.3$  \\
11904169 & $7.13\times10^{-1}$ & $6.73\times10^{-4} $ &  $42.3$ & 11904169 & $2.36$ & \nodata & \nodata  \\
11904171 & $2.22\times10^{-1}$ & $6.94\times10^{-4} $ &  $43.6$ & 11904171 & $7.35\times10^{-1}$ & $2.30\times10^{-3} $ &  $84.9$  \\
11904145 & $1.55\times10^{-1}$ & $4.04\times10^{-4} $ &  $25.4$ & 11904145 & $5.13\times10^{-1}$ & $1.34\times10^{-3} $ &  $49.5$  \\
11904150 & $1.10$ & \nodata & \nodata & 11904150 & $3.63$ & \nodata & \nodata  \\
11904152 & $1.95\times10^{-1}$ & $5.46\times10^{-4} $ &  $34.3$ & 11904152 & $6.46\times10^{-1}$ & $1.81\times10^{-3} $ &  $66.8$  \\
11904154 & $3.08\times10^{-1}$ & $6.51\times10^{-4} $ &  $40.9$ & 11904154 & $1.02$ & \nodata & \nodata  \\
11904155 & $1.56$ & \nodata & \nodata & 11904155 & $5.15$ & \nodata & \nodata  \\
11904158 & $3.09\times10^{-2}$ & $6.53\times10^{-4} $ &  $41.1$ & 11904158 & $1.02\times10^{-1}$ & $2.16\times10^{-3} $ &  $80.0$  \\
11904159 & $1.16\times10^{-1}$ & $5.63\times10^{-4} $ &  $35.4$ & 11904159 & $3.82\times10^{-1}$ & $1.86\times10^{-3} $ &  $69.0$  \\
11904160 & $2.51\times10^{-2}$ & $2.04\times10^{-4} $ &  $12.8$ & 11904160 & $8.31\times10^{-2}$ & $6.74\times10^{-4} $ &  $24.9$  \\
11904162 & $1.06\times10^{-1}$ & $4.09\times10^{-4} $ &  $25.7$ & 11904162 & $35.1\times10^{-1}$ & $1.35\times10^{-3} $ &  $50.1$  \\ [-1.5ex]
\enddata
\tablecomments{Transits on some companions can be ruled out because they require depth $>1$.}
\end{deluxetable}

\begin{table}[ht]%
\centering
\begin{tabular}{rrrr}
Distance &  Distance & $\Delta$J & J \\
(FWHM) & $(arcsec)$ & (mag) & (mag)\\ \hline
1 &  0.075 &  1.5 & 11.4 \\
2 &  0.150 &  3.5 & 13.4 \\
3 &  0.225 &  5.0 & 14.9 \\
4 &  0.300 &  5.0 & 14.9 \\
5 &  0.375 &  5.5 & 15.4 \\
6 &  0.450 &  5.5 & 15.4 \\
7 &  0.525 &  6.5 & 16.4 \\
8 &  0.600 &  7.0 & 16.9 \\
9 &  0.675 &  7.5 & 17.4 \\
40 &  3.000 &  9.5 & 19.4
\\
\end{tabular}
\caption{Palomar AO source sensitivity as a function of
distance from the primary target at $J$.\label{tab:paloAO}}
\end{table}

\begin{deluxetable}{rrrrrr}
\tablewidth{0pc}
\tablecaption{Relative Radial Velocity Measurements of \koi.\label{tab:velocities}}
\tablehead{
\colhead{HJD}                           &
\colhead{RV}                            &
\colhead{\ensuremath{\sigma_{\rm RV}}}  \\
\colhead{(-2450000)}                    &
\colhead{(\ms)}                         &
\colhead{(\ms)}                         &
}
\startdata
 5074.878 &    4.06 &    1.6  \\
 5075.773 &    6.83 &    1.5  \\
 5076.863 &    0.40 &    1.5  \\
 5077.923 &   -6.36 &    1.5  \\
 5078.922 &    1.76 &    1.7  \\
 5079.973 &   -2.41 &    1.7  \\
 5080.896 &    5.34 &    1.5  \\
 5081.969 &   -7.16 &    1.5  \\
 5082.848 &   -8.50 &    1.5  \\
 5083.761 &    3.27 &    1.4  \\
 5083.945 &   -2.27 &    1.6  \\
 5084.878 &    1.17 &    1.4  \\
 5106.890 &   -0.14 &    1.7  \\
 5169.725 &    1.58 &    0.9  \\
 5170.725 &   -3.00 &    1.1  \\
 5172.756 &    5.05 &    1.6  \\
 5173.721 &    4.90 &    1.1  \\
 5312.047 &   -2.68 &    1.6  \\
 5313.004 &   -5.29 &    1.5  \\
 5314.005 &   -9.82 &    1.6  \\
 5317.998 &    3.92 &    1.7  \\
 5318.121 &   -0.70 &    1.5  \\
 5319.027 &   -5.43 &    1.6  \\
 5320.063 &    1.44 &    1.5  \\
 5321.007 &    0.97 &    1.6  \\
 5321.969 &    1.61 &    1.5  \\
 5343.050 &   -4.01 &    1.5  \\
 5344.032 &   -4.93 &    1.5  \\
 5344.973 &   -4.08 &    1.3  \\
 5345.068 &   -2.08 &    1.7  \\
 5350.973 &    0.56 &    1.6  \\
 5351.988 &    0.88 &    1.4  \\
 5373.814 &   -0.19 &    1.6  \\
 5376.865 &   -0.38 &    1.4  \\
 5379.902 &    1.78 &    1.7  \\
 5403.898 &    0.66 &    1.4  \\
 5407.013 &   -3.80 &    1.5  \\
 5411.986 &   -3.83 &    1.4  \\
 5412.805 &   -4.36 &    1.2  \\
 5414.803 &   -1.20 &    1.5  \\
\enddata
\end{deluxetable}

\begin{deluxetable}{lcc}
\tabletypesize{\scriptsize}
\tablewidth{0pc}
\tablecaption{Star and planet parameters for the \starname\ system.\label{tab:SystemParams}}
\tablehead{\colhead{Parameter}	& 
\colhead{Value} 		& 
\colhead{Notes}}
\startdata
\sidehead{\em Transit and orbital parameters: \planetb}
Orbital period $P$ (days)			& \periodb		& A	\\
Midtransit time $E$ (BJD)			& \epochb		& A	\\
Scaled semimajor axis $a/\rstar$		& \scaledSemiMajb	& A	\\
Scaled planet radius \rpl/\rstar		& \scaledPlanetRadiusb	& A	\\
Impact parameter $b$                		& \impactb   		& A 	\\
Orbital inclination $i$ (deg)			& \inclinationb		& A	\\
Orbital semi-amplitude $K$ (\ms)		& \semiAmpb		& B	\\
Orbital eccentricity $e$			& \eccb 		& B	\\
Center-of-mass velocity $\gamma$ (\ms)		& \gammaVelb		& B	\\

\sidehead{\em Transit and orbital parameters: \koic}

Orbital period $P$ (days)                       & \periodc              & A     \\
Midtransit time $E$ (HJD)                       & \epochc               & A     \\
Scaled semimajor axis $a/\rstar$                & \scaledSemiMajc       & A     \\
Scaled planet radius \rpl/\rstar                & \scaledPlanetRadiusc  & A     \\
Impact parameter $b$                            & \impactc              & A     \\
Orbital inclination $i$ (deg)                   & \inclinationc         & A     \\

\sidehead{\em Observed stellar parameters}
Effective temperature \teff (K)			& \teffSME		& C 	\\
Spectroscopic gravity \logg (cgs)		& \loggSME		& C	\\
Metallicity \feh				& \fehSME		& C	\\
Projected rotation \vsini (\kms)		& \vsiniSME		& C	\\

\sidehead{\em Fundamental Stellar Properties}
Mass \mstar (\msun)				& \mstarKASC		& D	\\
Radius \rstar (\rsun)  				& \rstarKASC		& D	\\
Surface gravity \loggstar\ (cgs)		& \loggKASC		& D	\\
Luminosity \lstar\ (\lsun)			& \lumKASC		& D	\\
Absolute V magnitude $M_V$ (mag)		& \absMag   		& D	\\
Age (Gyr)					& \ageKASC		& D	\\
Distance (pc)					& \distance		& D	\\ 

\sidehead{\em Planetary parameters: \planetb}
Mass \mpl\ (\mearth)				& \mplanetb		& A,B,C,D	\\
Radius \rpl\ (\rearth)				& \rplanetb		& A,B,C,D	\\
Density \rhopl\ (\gcmc)				& \rhoplanetb		& A,B,C,D	\\
Surface gravity \loggpl\ (cgs)			& \loggplanetb		& A,B,C,D	\\
Orbital semimajor axis $a$ (AU)			& \semiMajb		& E		\\
Equilibrium temperature \teq\ (K)		& \teqb			& F		\\

\sidehead{\em Parameters for candidate: \koic}
Mass \mpl\ (\mearth)                              & $< 20$              & G       \\
Radius \rpl\ (\rearth)                            & \rplanetc             & A,D     \\
Orbital semimajor axis $a$ (AU)                 & \semiMajc		& E       \\
Equilibrium temperature \teq\ (K)               & \teqc			& F

\enddata
\tablecomments{
A: Based primarily on an analysis of the photometry,\\
B: Based on a joint analysis of the photometry and radial velocities,\\
C: Based on an analysis by D. Fischer of the Keck/HIRES template spectrum using SME \citep{Valenti96},\\
D: Based on asteroseismology analysis
E: Based on Newton's revised version of Kepler's Third Law and the results from D,\\
F: Calculated assuming a Bond albedo of 0.1 and complete redistribution of heat for reradiation.
G: Upper limit corresponding to three times the 68.3\% credible interval from MCMC mass distribution.
}
\end{deluxetable}

\begin{deluxetable}{ccccccc}
\tabletypesize{\scriptsize}
\tablewidth{0pc}
\tablecaption{Blend frequency estimate for \koib \ based on the frequencies of transiting giant planets.\label{tab:statistics1}}
\tablehead{
\colhead{Kp range} &
\colhead{$\Delta Kp$} &
\colhead{Stellar density} &
\colhead{$\rho_{\rm max}$} &
\colhead{Stars} &
\colhead{Transiting Jupiters} &
\colhead{Transiting Neptunes}
\\
\colhead{(mag)} &
\colhead{(mag)} &
\colhead{per sq.\ deg} &
\colhead{(\arcsec)} &
\colhead{($\times 10^6$)} &
\colhead{6--15\,$R_{\earth}$, $f_{\rm Jup}=0.11$\%} &
\colhead{3.8--6\,$R_{\earth}$, $f_{\rm Nep}=0.074$\%}
\\
\colhead{} &
\colhead{} &
\colhead{} &
\colhead{} &
\colhead{} &
\colhead{($\times 10^{-6}$)} &
\colhead{($\times 10^{-6}$)} \\
\colhead{(1)} &
\colhead{(2)} &
\colhead{(3)} &
\colhead{(4)} &
\colhead{(5)} &
\colhead{(6)} &
\colhead{(7)}
}
\startdata
11.0--11.5  &  0.5  & \nodata&\nodata & \nodata& \nodata & \nodata  \\
11.5--12.0  &  1.0  & \nodata&\nodata & \nodata& \nodata & \nodata  \\
12.0--12.5  &  1.5  & \nodata&\nodata & \nodata& \nodata & \nodata  \\
12.5--13.0  &  2.0  & \nodata&\nodata & \nodata& \nodata & \nodata  \\
13.0--13.5  &  2.5  &   48   & 0.12  &  0.168 & 0.0002  & 0.0001   \\
13.5--14.0  &  3.0  &   87   & 0.15  &  0.475 & 0.0005  & 0.0003   \\
14.0--14.5  &  3.5  &  106   & 0.18  &  0.833 & 0.0009  & 0.0006   \\
14.5--15.0  &  4.0  &  131   & 0.20  &  1.270 & 0.0014  & 0.0009   \\
15.0--15.5  &  4.5  &  189   & 0.22  &  2.217 & 0.0024  & 0.0016   \\
15.5--16.0  &  5.0  &  185   & 0.25  &  2.803 & 0.0031  & 0.0020   \\
16.0--16.5  &  5.5  & \nodata&\nodata & \nodata& \nodata & \nodata  \\
16.5--17.0  &  6.0  & \nodata&\nodata & \nodata& \nodata & \nodata  \\
\noalign{\vskip 6pt}
\multicolumn{2}{c}{Totals} & 746  &\nodata & 7.766 & 0.0085  & 0.0055   \\
\noalign{\vskip 4pt}
\hline
\noalign{\vskip 4pt}
\multicolumn{7}{c}{Blend frequency (BF) = $(0.0085 + 0.0055)\times 10^{-6}= 1.4 \times 10^{-8}$} \\
\enddata

\tablecomments{Column 1: magnitude bins; column 2: magnitude
  difference $\Delta Kp$ compared to the primary, taken at the upper
  edge of each bin; column 3: density of stars from the Besan\c{c}on
  models; column 4: angular separation at which stars in the
  corresponding magnitude bin would go undetected in our imaging
  observations; column 5: number of stars in a circle of radius
  $\rho_{\rm max}$ around \koi, restricted also by mass according
  to the contours in Figure~\ref{fig:blender1}; column 6: number of
  expected transiting Jupiters as contaminants; column 7: number of
  expected transiting Neptunes as contaminants.}

\end{deluxetable}


\clearpage

\clearpage

\clearpage

\clearpage

\clearpage

\clearpage

\clearpage

\clearpage

\clearpage

\clearpage

\clearpage

\clearpage

\clearpage

\clearpage

\clearpage

\clearpage

\begin{thebibliography}{}

\bibitem[Agol et al.(2005)]{agol}
Agol, E., Steffen, J., Sari, R., Clarkson, W. 2005, \mnras, 359, 567 

\bibitem[Argabright et al.(2008)]{argabright08}
Argabright, V.S., VanCleve, J.E., Bachtell, E.E., Hegge, M.J., McArthur, S.P., Dumont, F.C., Rudeen, A.C., Pullen, J.L., Teusch, D.A., Tennant, D.S. and Atcheson, P.D. 2008, Space Telescopes and Instrumentation 2008: Optical, Infrared, and Millimeter, eds. Oschmann, J.M., Jr., de Graauw, M.W.M., MacEwen, H.A., Proceedings of the SPIE, 7010, 70102

\bibitem[Batalha et al.(2010)]{batalha_tm}
Batalha, N.~M., Borucki, W.~J., Koch, D.~G., Bryson, S.~T., Haas, M.~R., Brown, T.~M., Caldwell, D.~A., Hall, J.~R., Gilliland, R.~L., Latham, D.~W., Meibom, S., Monet, D.~G. 2010, \apjl, 713, 109

\bibitem[Batalha et al.(2010)]{batalha_fp}
Batalha, N.M., Rowe, J.F., Gilliland, R.L., Jenkins, J.J., Caldwell, D., Borucki, W.J., Koch, D.G., Lissauer, J.J., Dunham, E.W., Gautier, T.N., Howell, S.B., Latham, D.W., Marcy, G.W., Prsa, A. 2010, \apjl, 713, 103

\bibitem[Borucki et al.(2009)]{borucki:09}
Borucki, W.~J. et al. 2009, Science, 325, 709

\bibitem[Borucki et al.(2010a)]{kepler4b}
 Borucki, W.~J. et al. 2010, \apjl, 713, 126

\bibitem[Borucki et al.(2010b)]{Borucki:10}
 Borucki, W.\ J.\ et al.\ 2010, \apj, submitted (arXiv:1006.2779)

\bibitem[Bruntt et al.(2010)]{bruntt}
Bruntt, H., et al. 2010, \aap, 519, 51 

\bibitem[Bryson et al.(2010)]{Bryson:10}
Bryson, S., Tenenbaum, P., Jenkins, J.M., Chandrasekharan, H., Klaus, T., Caldwell, D.A., Gilliland, R.L., Haas, M.R., Dotson, J.L., Koch, D.G., Borucki, W.J. 2010, \apjl, 713, 97

\bibitem[Caldwell et al.(2010)]{caldwellSPIE}
Caldwell, D.A., van Cleve, J.E., Jenkins, J.M., Argabright, V.S., Kolodziejczak, J.J., Dunham, E.W., Geary, J.C., Tenenbaum, P., Chandrasekaran, H., Li, J., Wu, H., von Wilpert, J. 2010, Space Telescopes and Instrumentation 2010: Optical, Infrared, and Millimeter Wave, eds. Oschmann, J.M., Jr.; Clampin, M.C., MacEwen, H.A., Proceedings of the SPIE, 7731, 773117

\bibitem[Charbonneau et al.(2009)]{charbonneau}
Charbonneau, D. et al., 2009, Nature, 462, 891

\bibitem[Christensen-Dalsgaard(2004)]{Christ2004}
Christensen-Dalsgaard, J., 2004.
[Physics of solar-like oscillations].
{\it Solar Phys.}, {\bf 220}, 137 -- 168.

\bibitem[Christensen-Dalsgaard(2008a)]{Christ2008a}
Christensen-Dalsgaard, J., 2008a.
[ASTEC -- the Aarhus STellar Evolution Code].
{\it Astrophys. Space Sci.},  {\bf 316}, 13 -- 24.

\bibitem[Christensen-Dalsgaard(2008b)]{Christ2008b}
Christensen-Dalsgaard, J., 2008b.
[ADIPLS -- the Aarhus adiabatic pulsation package].
{\it Astrophys. Space Sci.},  {\bf 316}, 113 -- 120.

\bibitem[Christensen-Dalsgaard {\it et~al.\/}(2008)]{Christetal2008}
Christensen-Dalsgaard, J., Arentoft, T., Brown, T. M., Gilliland, R. L.,
Kjeldsen, H., Borucki, W. J. \& Koch, D., 2008.
[The {\it Kepler\/} Asteroseismic Investigation].
In {\it Proc. HELAS II International Conference: Helioseismology,
Asteroseismology and the MHD Connections, G\"ottingen, August 2007},
eds L. Gizon \& M. Roth,
{\it J. Phys.: Conf. Ser.}, {\bf 118}, 012039(1 -- 10).

\bibitem[Christensen-Dalsgaard {\it et~al.\/}(2010)]{Christ2010}
Christensen-Dalsgaard, J., Kjeldsen, H., Brown, T.~M., Gilliland, R.~L.,
Arentoft, T., Frandsen, S., Quirion, P.-O., Borucki, W.~J., Koch, D. \&
Jenkins, J.~M., 2010.
[Asteroseismic investigation of known planet hosts in the {\it Kepler} field].
{\it Astrophys. J.}, {\bf 713}, L164 -- L168.

\bibitem[Dunham et al.(2010)]{kepler6b}
Dunham, E.~W. et al. 2010, \apjl, 713, 136

\bibitem[Ferraz-Mello et al.(2010)]{ferraz}
Ferraz-Mello, S., Tadeu dos Santos, M., Beauge, C., Michtchenko, T.A., Rodriguez, A. 2010, arXiv:1011.2144

\bibitem[Ford(2005)]{ford:05}
Ford, E.~B. 2005, \aj, 129, 1706

\bibitem[Ford(2006)]{ford:06}
Ford, E.~B. 2006, \apj, 642, 505

\bibitem[Fortney, Marley \& Barnes(2007)]{fortney}
Fortney, J., Marley, M., Barnes, J. 2007, \apj, 659, 1661

\bibitem[French et al.(2009)]{french}
French, M. et al. 2009, Phys. Ref. B, 79, 054107

\bibitem[Gilliland et al.(2010)]{gilliland:10}
Gilliland, R. L., et al. 2010, \apjl, 713, L160.

\bibitem[Girardi et al.(2008)]{girardi}
Girardi L., Dalcanton J., Williams B., de Jong R., Gallart C., Monelli M., Groenewegen M.A.T., Holtzman J.A., Olsen K.A.G., Seth A.C., Weisz D.R. 2008, \pasp, 120, 583

\bibitem[Grasset, Schneider, \& Sotin(2009)]{grasset}
Grasset, O., Schneider, J., Sotin, C. 2009, \apj, 693, 722

\bibitem[Grec {\it et~al.\/}(1983)]{Grec1983}
Grec, G., Fossat, E. \& Pomerantz, M., 1983.
[Full-disk observations of solar oscillations from the geographic South
Pole: latest results].
{\it Solar Phys.}, {\bf 82}, 55 -- 66.

\bibitem[Grevesse \& Noels(1993)]{Greves1993}
Grevesse, N. \& Noels, A., 1993.
[Cosmic abundances of the elements].
In {\it Origin and evolution of the Elements},
eds N. Prantzos, E. Vangioni-Flam \& M. Cass\'e
(Cambridge: Cambridge Univ. Press), 15 -- 25.

\bibitem[Hatzes et al.(2010)]{hatzes}
Hatzes, A.P. et al. 2010 \aap, 520, 93

\bibitem[Hayward et al.(2001)]{hayward2001} Hayward, T.~L., Brandl, B.,
Pirger, B., Blacken, C., Gull, G.~E., Schoenwald, J., \& Houck, J.~R.\
2001, \pasp, 113, 105

\bibitem[Holman et al.(2010)]{kepler9bc}
Holman, M.~J. et al. 2010, Science, 330, 51

\bibitem[Howell et al.(2010)]{howell10} Howell, S. B. et al. 2010, in preparation

\bibitem[Huber {\it et~al.\/}(2009)]{Huber2009}
Huber, D., Stello, D., Bedding, T. R., Chaplin, W. J., Arentoft, T.,
Quirion, P.-O. \& Kjeldsen, H., 2009.
[Automated extraction of oscillation parameters for Kepler observations
of solar-type stars].
{\it Comm. in Asteroseismology}, {\bf 160}, 74 -- 91.

\bibitem[Isaacson \& Fischer(2010)]{Isaacson10} Isaacson,H., \& Fischer,
D. A.\ 2010, ApJ, 725, 875

\bibitem[Jenkins et al.(2010a)]{socpipeline} 
Jenkins, J.~M., et al. 2010,  \apjl {\bf 713(2)}, L87.

\bibitem[Jenkins et al.(2010b)]{jenkins:10}
Jenkins, J.~M., et al. 2010, \apjl, 713, L120.

\bibitem[Jenkins et al.(2010c)]{kepler8b}
Jenkins, J.~M. et al. 2010, 724, 1108  

\bibitem[Jenkins et al.(2010d)]{jenkinsSPIE} 
Jenkins, J.~M., Chandrasekaran, H., McCauliff, S.~D., Caldwell, D.~A., Tenenbaum, P., Li, J., Klaus, T.~C., Cote, M.~T., Middour, C. 2010, \procspie, 7740, 77400D

\bibitem[Johnson et al.(2009)]{Johnson09} Johnson, J.~A., Winn,
J.~N., Albrecht, S., Howard, A.~W., Marcy, G.~W.,
\& Gazak, J.~Z.\ 2009, \pasp, 121, 1104

\bibitem[Koch et al.(2010)]{koch10} Koch, D. G. et al. 2010, \apjl, 713,79

\bibitem[Koch et al.(2010)]{kepler5b} Koch, D. G. et al. 2010, \apjl, 713, 131

\bibitem[Latham et al.(2005)]{kicLatham} Latham, D. W., Brown, T. M., Monet, D. G., Everett, M., Esquerdo, G. A., Hergenrother, C. W. 2005, \baas, 37, 1340

\bibitem[Latham et al.(2010)]{kepler7b}
Latham, D.~W. et al. 2010, \apjl, 713, 140

\bibitem[Leger et al.(2009)]{leger}
Leger, A et al. 2009 \aap,506,287

\bibitem[Mandel \& Agol(2002)]{man02} 
Mandel, K. \& Agol, E., 2002, \apj, 580, 171

\bibitem[Mandushev et al.(2005)]{mandushev}
Mandushev, G., et al. 2005, \apj, 621, 1061

\bibitem[Marcus et al.(2010)]{marcus}
Marcus, R., Sasselov, D., Hernquist, L., Stewart, S. 2010, \apjl, 712, 73

\bibitem[Marcy et~al. (2008)]{Marcy08}Marcy, G.~W., et al.\ 2008,
  Physica Scripta Volume T, 130, 014001

\bibitem[Marigo et al.(2008)]{Marigo:08}
 Marigo, P., Girardi, L., Bressan, A., Groenewegen, M.\ A.\ T., Silva,
 L., \& Granato, G.\ L. 2008, \aap, 482, 883

\bibitem[Mazeh(2008)]{Mazeh:08} Mazeh, T. 2008, in Tidal Effects in
Stars, Planets and Disks, EAS Publications Series, eds.\ M.-J.\ Goupil
\& J.-P.\ Zahn (EDP Sciences), Vol.\ 29, p. 1

\bibitem[Nettelmann et al.(2010)]{nettelmann}
Nettelmann, N., Fortney, J.~J., Kramm, U., Redmer, R. 2010, arXiv:1010.0277

\bibitem[Pfahl et al.(2008)]{pfahl}
Pfahl, E., Arras, P. \& Paxton, B. 2008, \apj, 679, 783

\bibitem[Pont, Aigrain, \& Zucker(2010)]{pont}
Pont, F., Aigrain, S., Zucker, S. 2010, arXiv:1008.3859

\bibitem[Prsa(2010)]{limbdarkening}
Prsa, A. 2010, \url{http://astro4.ast.villanova.edu/aprsa/}

\bibitem[Queloz et al.(2001)]{queloz:01}
Queloz, D., et al. 2001, \aap, 379, 279

\bibitem[Queloz et al.(2009)]{queloz}
Queloz, D. et al. 2009, \aap, 506, 303

\bibitem[Ragozzine \& Holman(2010)]{ragozzine}
Ragozzine, D., Holman, M.~J. 2010, arXiv:1006.3727

\bibitem[Robin et al.(2003)]{Robin:03}
 Robin, A.\ C., Reyl\'e, C., Derri\'ere, S., \& Picaud, S. 2003, \aap,
 409, 523

\bibitem[Rogers \& Seager(2010)]{rogers}
Rogers, L.A. \& Seager, S. 2010, \apj, 712, 974

\bibitem[Seager \& Mallen-Ornelas(2003)]{seager:03}
Seager, S. \& Mall\'en-Ornelas, G 2003, \apj, 585, 1038

\bibitem[Seager \& Deming(2009)]{seager:09}
Seager, S. \& Deming, D. 2009, \apj, 703, 1884

\bibitem[Seager, Whitney, \& Sasselov(2000)]{seager:00}
Seager, S., Whitney, B., Sasselov, D. 2000, \apj,540,504

\bibitem[Seager et al.(2007)]{seager:07}
Seager, S. et al. 2007, \apj, 669, 1279

\bibitem[Steffen \& Agol(2005)]{steffen}
Steffen, J.~H. \& Agol, E. 2005, \mnras, 364, L96

\bibitem[Steffen et al.(2010)]{multis}
Steffen, J.H. et al. 2010, ApJ, 725, 1226

\bibitem[Tassoul(1980)]{Tassou1980}
Tassoul, M., 1980.
[Asymptotic approximations for stellar nonradial pulsations].
{\it Astrophys. J. Suppl.}, {\bf 43}, 469 -- 490.

\bibitem[Torres et al.(2004)]{Torres:04}
 Torres, G., Konacki, M., Sasselov, D.~D., \& Jha, S. 2004, \apj,
614, 979

\bibitem[Torres et al.(2005)]{Torres:05}
Torres, G., Konacki, M., Sasselov, D. D., Jha, S. 2005, \apj, 619, 558

\bibitem[Torres et al.(2010)]{Torres:10}
Torres, G.\ et al.\ 2010, \apj, 727, 24 

\bibitem[Troy et al.(2000)]{troy2000} Troy, M., et al.\ 2000, \procspie,
4007, 31

\bibitem[Twicken et al.(2010a)]{pa}
Twicken, J.~D., Clarke, B.~D., Bryson, S.~T., Tenenbaum, P., Wu, H., Jenkins, J.~M., Girouard, F., Klaus, T.~C.
2010, \procspie, 7740, 774023

\bibitem[Twicken et al.(2010b)]{pdc}
Twicken, J.~D., Chandrasekaran, H., Jenkins, J.~M., Gunter, J.~P., Girouard, F., Klaus, T.~C.
2010, \procspie, 7740, 77401U

\bibitem[Valencia et al.(2009)]{valencia:09}
Valencia, D.; Ikoma, M.; Guillot, T., Nettelmann, N. 2009, \aap, 516, id.A20

\bibitem[Valencia, O'Connell, \& Sasselov(2006)]{valencia:06}
Valencia, D., O'Connell, R., Sasselov, D. 2006, Icarus, 181, 545

\bibitem[Valencia, O'Connell, \& Sasselov(2007)]{valencia:07}
Valencia, D., Sasselov, D., O'Connell, R. 2007, \apj, 665, 1413

\bibitem[Valenti \& Piskunov(1996)]{Valenti96} Valenti, J.~A., \&
  Piskunov, N. \ 1996, \aaps, 118, 595

\bibitem[Valenti \& Fischer(2005)]{Valenti05} Valenti, J.~A.,
\& Fischer, D.~A.\ 2005, \apjs, 159, 141

\bibitem[Van Cleve \& Caldwell(2009)]{vancleve}  
Van Cleve, J., \& D. A. Caldwell 2009, Kepler Instrument Handbook, KSCI 19033-001, (Moffett Field, CA: NASA Ames Research Center) \url{http://archive.stsci.edu/kepler}

\bibitem[Vandakurov(1967)]{Vandak1967}
Vandakurov, Yu. V., 1967.
[The frequency distribution of stellar oscillations].
{\it Astron. Zh.}, {\bf 44}, 786 -- 797
(English translation: {\it Soviet Astronomy AJ}, {\bf 11}, 630 -- 638).

\bibitem[van Kerkwijk et al.(2010)]{dopplerboosting}
van Kerkwijk, M.~H., Rappaport, S,~A., Breton, R.~P., Justham, S., Podsiadlowski, P., Han, Z. 2010, \apj, 715, 51

\bibitem[{Vogt} {et~al.}(1994)]{vogt94} Vogt, S.~S., et al.\ 1994,
\procspie, 2198, 362

\bibitem[Welsh et al.(2010)]{welsh}
Welsh, W.F., Orosz, J.A., Seager, S., Fortney, J., Jenkins, J., Rowe, J.F., Koch, D., Borucki, W.J. 2010, \apjl, 713,145

\bibitem[Wu et al.(2010)]{dv}
Wu, H., Twicken, J.D., Penenbaum, P., Clarke, B.D., Li, J., Quintana, E.V., Allen, C., Chandrasekaran, H., Jenkins, J.M, Caldwell, D.A., Wohler, B., Girouard, F., McCauliff, S., Cote, M.T., Klaus, T.C. 2010, 
Software and Cyberinfrastructure for Astronomy, Proceedings of the SPIE, eds. Radziwill, N.M.; Bridger, A., 
7740, 42

\bibitem[Zeng \& Sasselov(2010)]{zeng}
Zeng, L. \& Sasselov, D. 2011, \apj, submitted

\end{thebibliography}
\end{document}